\documentclass[journal,10pt]{IEEEtran}
 \usepackage{booktabs}
\usepackage{multirow}
\usepackage[table,xcdraw]{xcolor}
\normalsize

\ifCLASSINFOpdf
 \usepackage[pdftex]{graphicx}
\else
 \usepackage[dvips]{graphicx}
\fi

\DeclareGraphicsExtensions{.eps}
\usepackage[cmex10]{amsmath}
\usepackage{bm}
\usepackage{upgreek}
\usepackage{epsfig}
\usepackage[noadjust]{cite}
\usepackage{latexsym}
\usepackage{multirow}
\usepackage{epstopdf}
\usepackage{color}  
\usepackage{comment}
\usepackage{tabularx} 
\usepackage{amsfonts,amssymb,dsfont}
\graphicspath{{./Figures/}}
\usepackage{color}
\usepackage{bbm}
\usepackage{mathtools}
\usepackage{mathabx}
\usepackage{siunitx}
\usepackage{algorithm}
\usepackage{algorithmicx}
\usepackage{algpseudocode}
\usepackage{url}

\usepackage{cite}
\usepackage[tight,footnotesize]{subfigure}
\usepackage{balance}

\usepackage{color, soul}

\newcommand{\ch}[1]{\textcolor{red}{{#1}}}

\newcommand{\response}[1]{\textcolor{red}{{#1}}}
\usepackage{textcomp}
\usepackage{caption}
\DeclareCaptionLabelSeparator{otherLine}{\\}
\DeclareCaptionLabelSeparator{periodTwoSpace}{.\ ~}
\usepackage{url}
\usepackage[pagebackref=true,breaklinks=true,letterpaper=true,colorlinks,bookmarks=false]{hyperref}
\usepackage[hyphenbreaks]{breakurl}

\usepackage{siunitx}
\usepackage{tikz,array}
\usetikzlibrary{trees}

\DeclareSIUnit\bps{bps}
\DeclareSIUnit\Torr{Torr}
\DeclareSIUnit\torr{Torr}
\DeclareSIUnit\sample{Sa}

\usepackage{cite}
\usepackage{pifont}
\newcommand{\cmark}{\ding{51}}%
\newcommand{\xmark}{\ding{55}}%
\usepackage[table,xcdraw]{xcolor}
\usepackage{longtable}

\begin{document}

\title{\huge Terahertz Band Communication:
An Old Problem Revisited and Research Directions for the Next Decade
\\ \textit{(Invited Paper)}
}
\author{Ian F. Akyildiz, \textit{Life Fellow, IEEE}, Chong~Han, \textit{Member, IEEE,} Zhifeng~Hu,\\
Shuai Nie, \textit{Member, IEEE,} and Josep M. Jornet, \textit{Senior Member, IEEE}


\thanks{Ian F. Akyildiz is with Truva Inc., Alpharetta, GA 30022, USA (email: ian@truvainc.com).
\par Chong Han and Zhifeng Hu are with the Terahertz Wireless Communications (TWC) Laboratory, Shanghai Jiao Tong University, Shanghai 200240, China (email: \{chong.han, zhifeng.hu\}@sjtu.edu.cn).
\par Shuai Nie is with the School of Computing, University of Nebraska-Lincoln, Lincoln, NE 68588, USA (email: shuainie@unl.edu).

\par Josep M. Jornet is with the Ultrabroadband Nanonetworking Laboratory, Institute for the Wireless Internet of Things, Department of Electrical and Computer Engineering, Northeastern University, Boston, MA 02215, USA (email: jmjornet@northeastern.edu).
}
}

%
%

\markboth{IEEE Transactions on Communications, 2022}
{}
\maketitle

\thispagestyle{empty} 

\begin{abstract}
Terahertz (THz) band communications are envisioned as a key technology for 6G and Beyond.
As a fundamental wireless infrastructure, THz communication can boost abundant promising applications.
In 2014, our team published two comprehensive roadmaps for
the development and progress of  THz communication networks~\cite{Akyildiz,teranets}, which helped the research community to start research on this subject afterwards. The topic of THz communications became very important and appealing to the research community due to 6G wireless systems design and development in recent years. Many papers are getting published covering different aspects of wireless systems using the THz band. With this paper, our aim is looking back to the last decade and revisiting the old problems and pointing out what has been achieved in the research community so far. Furthermore,  {in this paper, open challenges and new research directions still to be investigated for the THz band communication systems are presented, by covering diverse {topics ranging from devices, channel behavior, communication and networking, to physical testbeds and demonstration systems.}} The key aspects presented in this paper will enable THz communications as a pillar of 6G and Beyond wireless systems in the next decade.
\end{abstract}


\section{Introduction}

\begin{figure*}
	\centering
	\includegraphics[width=\textwidth]{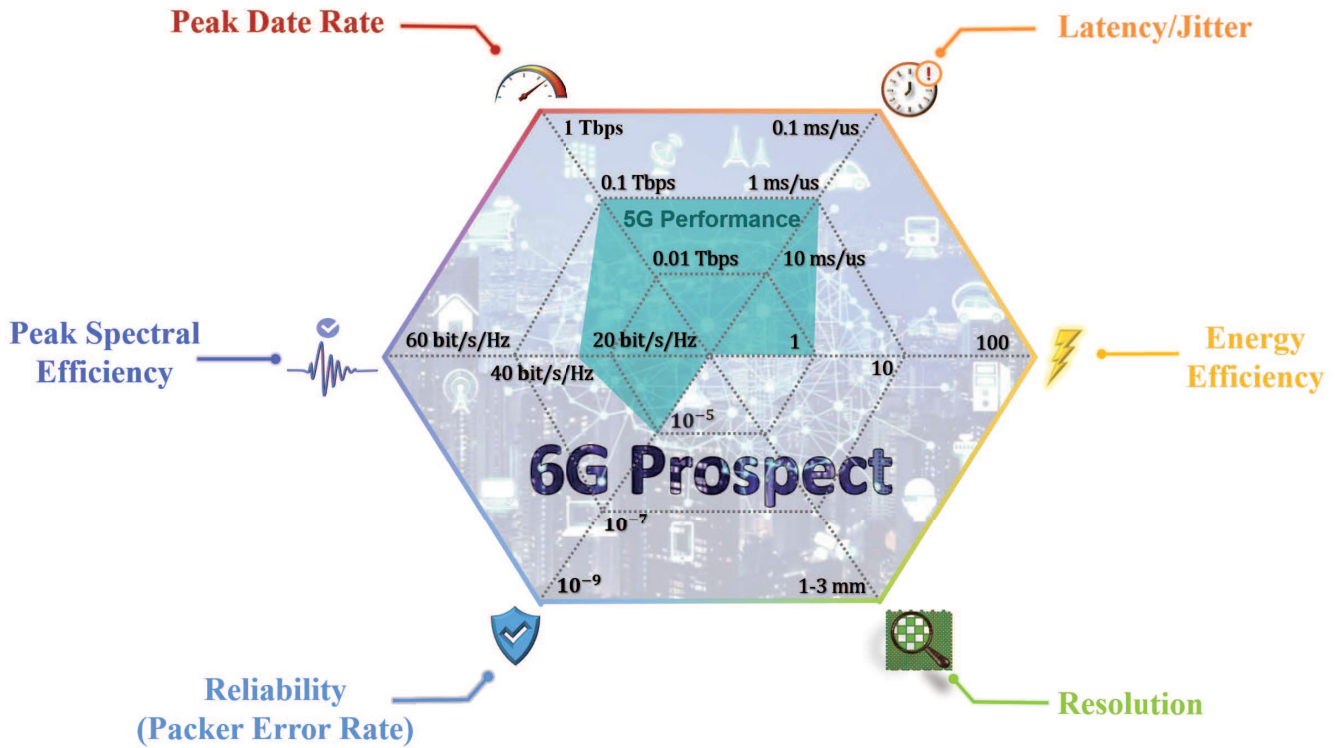}
	\caption{Performance objectives for 6G.}\label{fig:6G_views}
\end{figure*}

Terahertz (THz) Band (0.1--10~THz) communications is envisioned as one of the key enabling technologies to satisfy the exponential growth of data traffic volume accompanied by the demand for higher data rates and better coverage in 6G and Beyond wireless systems.
As shown in Fig.~\ref{fig:6G_views}, 6G wireless systems are expected to support peak data rates of 1~Tbps, where the peak spectral efficiency is expected to be 60 bps/Hz, end-to-end reliability in terms of packet error rates of $10^{-9}$, 
and end-to-end latency of 0.1 ms. 
Furthermore, the energy efficiency will be improved 100 times better than in 5G, as well as 1 to 3~mm sensing resolutions in the next generation Internet of Things with ultra-massive number of devices, e.g., on a scale of tens of billions ~\cite{akyildiz20206g,Akyildiz,teranets,Elayan2020THz,rappaport2019wireless,Boulogeorgo2018THz}.


Millimeter-wave (mm-wave) communications (30--300~GHz) under 100 GHz have been officially adopted in recent 5G cellular systems.
While the trend for higher carrier frequencies is apparent, it is still difficult for mm-wave systems to support Tbps data rates as well as the above mentioned strict Quality of Service requirements. 
Limited by the total consecutive available bandwidth of less than 10~GHz in the mm-wave systems under 100 GHz, the spectrum efficiency is required to reach 100~bps/Hz, which is unprecedentedly challenging even with advanced physical-layer techniques.
THz band reveals its potential as one of key wireless technologies to fulfill the future demands for 6G wireless systems, thanks to its four strengths: 
1) from tens and up to hundreds of GHz of contiguous bandwidth, 
2) picosecond-level symbol duration, 
3) integration of thousands of sub-millimeter-long antennas, 
4) ease of coexistence with other regulated and standardized spectrum.
The THz band has traditionally been one of the least explored frequency bands in the electromagnetic (EM) spectrum, mostly due to the lack of efficient and practical THz transceivers and antennas. 
 {Nevertheless, to expedite the way of fulfilling this gap, practical THz communication systems are being enabled by the major progress in the last ten years~\cite{song2021TTHz,crowe2017terahertz,Huq2019THz}.}

\begin{figure*}
    \centering
    \includegraphics[width=\textwidth]{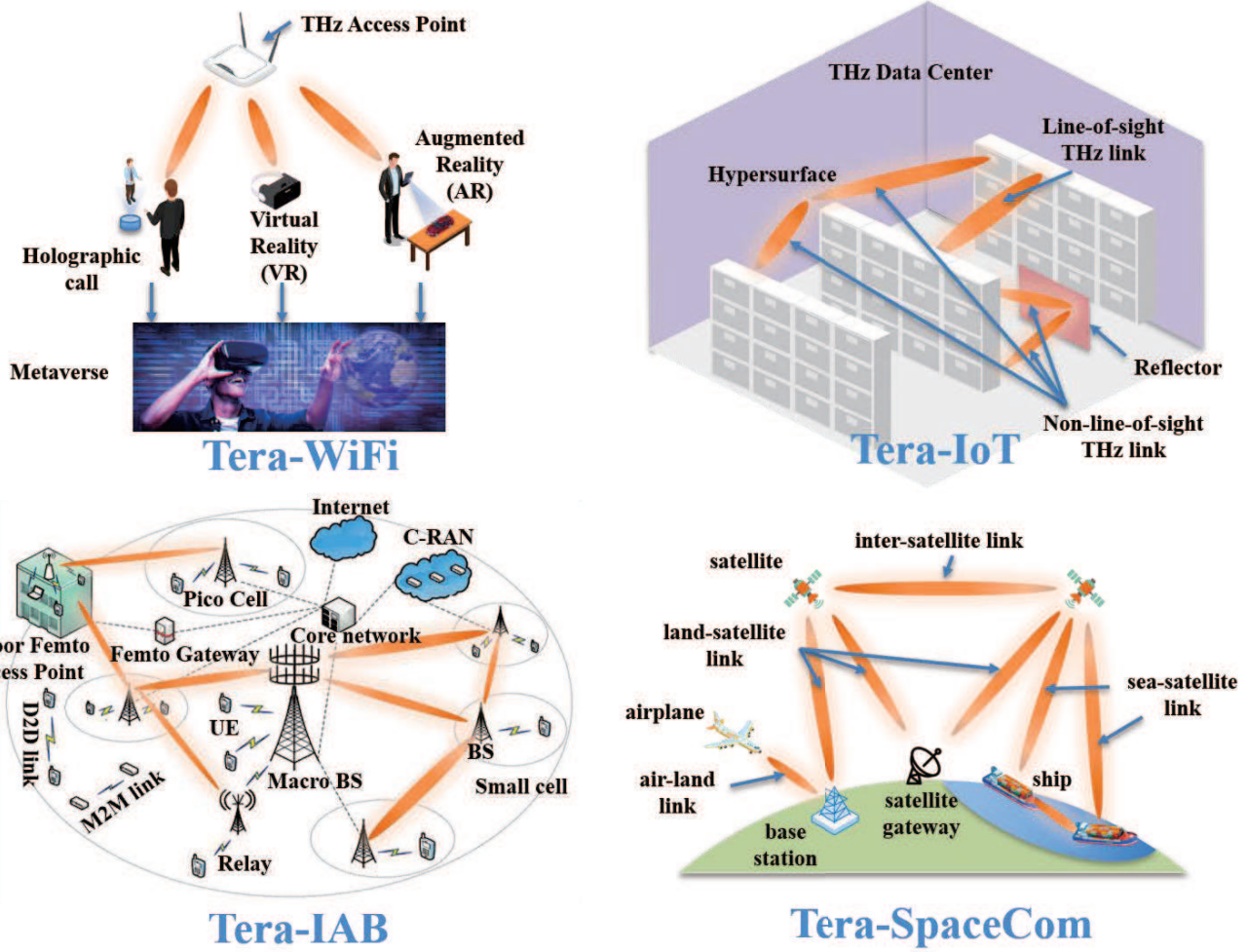}
    \caption{6G applications enabled by THz communications, including Tbps WLAN system (\textit{Tera-WiFi}), Tbps Internet-of-Things (\textit{Tera-IoT}) in wireless data center, Tbps integrated access and backhaul (\textit{Tera-IAB}) wireless networks, and ultra-broadband THz space communications (\textit{Tera-SpaceCom}).
    }
    \label{fig:applications}
\end{figure*}

\begin{figure*}
\centering
	\subfigure[]{ \includegraphics[width=0.47\textwidth]{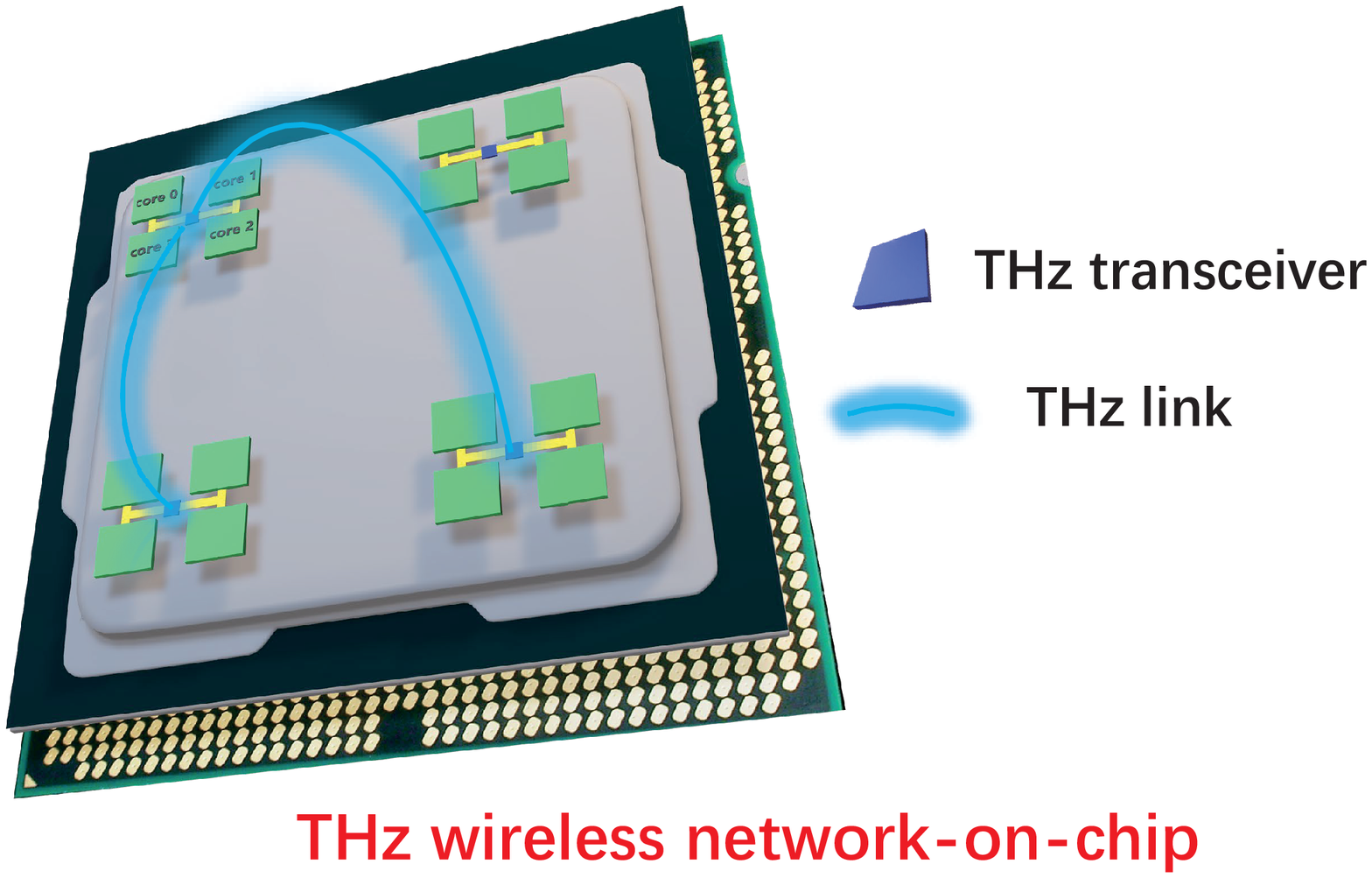} \label{fig:winoc}}
	\subfigure[]{ \includegraphics[width=0.47\textwidth]{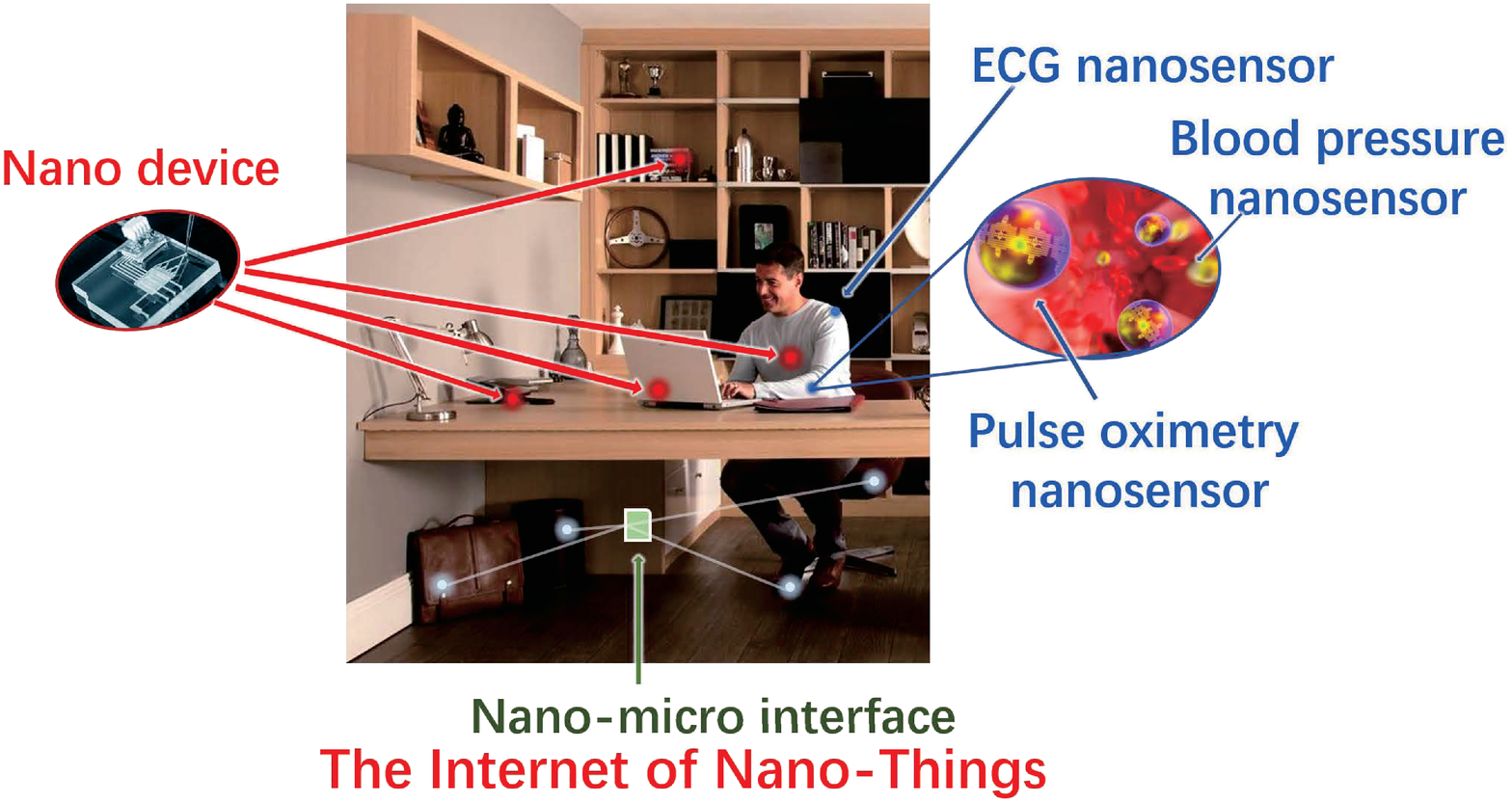} \label{fig:iont}}
\caption{Nanoscale applications enabled by THz-band communications, including (a) wireless networks-on-chip communications (WiNoC), and (b) the Internet of Nano-Things (IoNT).}
        \label{fig:nanoapp}
\end{figure*}

\begin{figure*}
	\centering
	\includegraphics[width=0.95\textwidth]{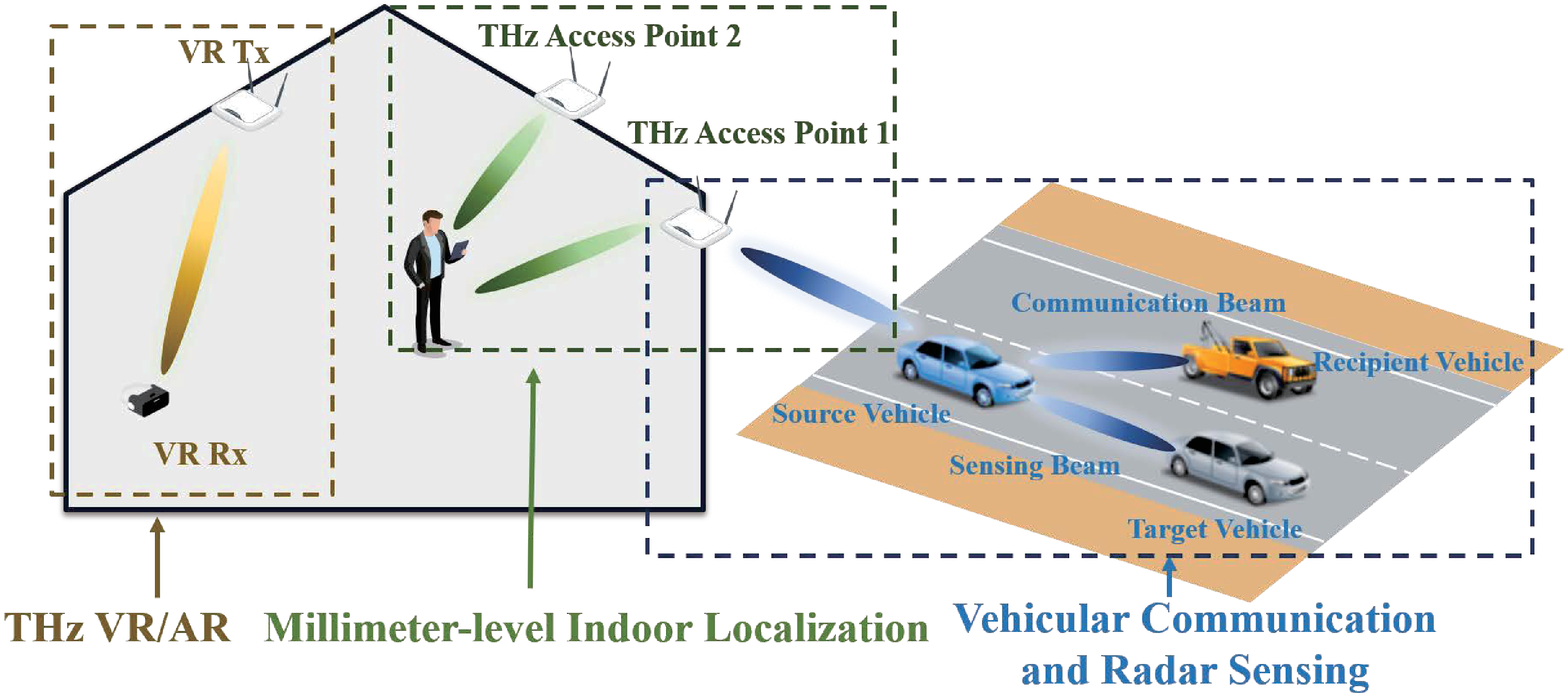}
	\caption{THz integrated sensing and communication applications.}
	\label{fig:jcs}
\end{figure*}

With the multi-GHz bandwidth, the THz spectrum is favored as a candidate to resolve the spectrum scarcity problem and tremendously enhance the capacity of current wireless systems~\cite{Petrov2018last}.
In addition to the promised Tbps-level links for cellular systems, THz-band spectrum can also be utilized in many use cases, such as Tbps WLAN system (\textit{Tera-WiFi}), Tbps Internet-of-Things (\textit{Tera-IoT}) in wireless data centers, Tbps integrated access backhaul (\textit{Tera-IAB}) wireless networks, and ultra-broadband THz space communications (\textit{Tera-SpaceCom}), as illustrated in Fig.~\ref{fig:applications}.
Besides macro/micro-scale applications, motivated by the state-of-the-art nanoscale transceivers and antennas that intrinsically operate in the THz band, further use cases include wireless networks-on-chip communications (WiNoC) as well as wireless connections in networks of nanomachines such as in Internet of NanoThings (IoNT)~\cite{nanothings}, shown in Fig.~\ref{fig:nanoapp} where 
these nano-devices can be the building block for the intra-body communications~\cite{Abbasi2017Cooperative, Elayan2018End} to provide advanced healthcare services.
To further integrate sensing and communication (ISAC), also known as joint communication and sensing (JCS) capabilities, the promising applications, containing THz virtual/augmented reality (VR/AR), vehicular communication and radar sensing, and millimeter-level indoor localization, are shown in Fig.~\ref{fig:jcs}~\cite{wu2021ISCI}.
Moreover, powerful deep learning (DL) schemes can further enable THz holographic, haptic and telepresence communication, or metaverse, to offer a more intuitive visual and aesthetic experience for entertainment, academic research, and so on.
 {We envision various applications by using the THz spectrum, without specifying the exact bands. These applications are supported by using the THz band, whereas it would be hard to classify the suitable bands for each application. Although the lower THz band (e.g., less than 400 GHz) is mostly used and studied till date, this should not imply the trend of further moving from lower to upper THz band in the future, with the steady progress of standardization and device technologies.}

The road to close the THz gap and realize THz communications needs to be paved by the communication community jointly.
In this direction, in 2019, the FCC (Federal Communication Commission) created a new category of experimental licenses and allocated over 20~GHz of unlicensed spectrum between 95~GHz and 3~THz, to facilitate the testing of 6G and Beyond technologies, including those funded by multiple  {National Science Foundation (NSF)} and Department of Defense (DoD) grants as well as industry driven efforts. 
Global research activities are ongoing as well, including Europe Horizon 2020 and key programs funded by Chinese Ministry of Science and Technology~\cite{eu6g}.
In 2014, we published two comprehensive roadmap papers for the development of THz communication networks~\cite{Akyildiz,teranets}, which helped the research community to pay attention to this upcoming important subject. In particular, many people started research on this topic and the subject became like a wildfire with hundreds of papers getting published in recent years. With this paper, we want to look back at the last decade and revisit the old problems and point out what has been achieved in the research community. 
Furthermore, we look into the next decade and
point out all the remaining research challenges so that THz Band
wireless communication can be fully utilized. More specifically, our main contributions are
summarized as follows.
\begin{itemize}

\item We present the development of pathways for THz devices, including analog front-ends, reconfigurable antenna systems and ultra-broadband digital back ends.

\item We provide a thorough inspection on THz channels, containing measurement methods, modeling schemes, unique propagation properties, and link budget analysis.

\item In light of the device and channel characteristics, we conduct a detailed review of the THz physical-layer technologies, including modulation, coding, dynamic hybrid beamforming, beam estimation and tracking, synchronization, localization, as well as physical layer security.

\item Looking up to higher-layer networking protocols, we inspect the medium access control layer, interference and coverage, multi-hop communications, together with the routing and scheduling approaches in THz wireless communication systems.

\item We also present the advancement of THz experimental and
simulation testbeds.

\item We discuss the open challenges and present potential future research directions from all the aforementioned aspects of THz communications for the next decade.
\end{itemize}

\begin{figure*}
	\centering
	\includegraphics[width=1\textwidth]{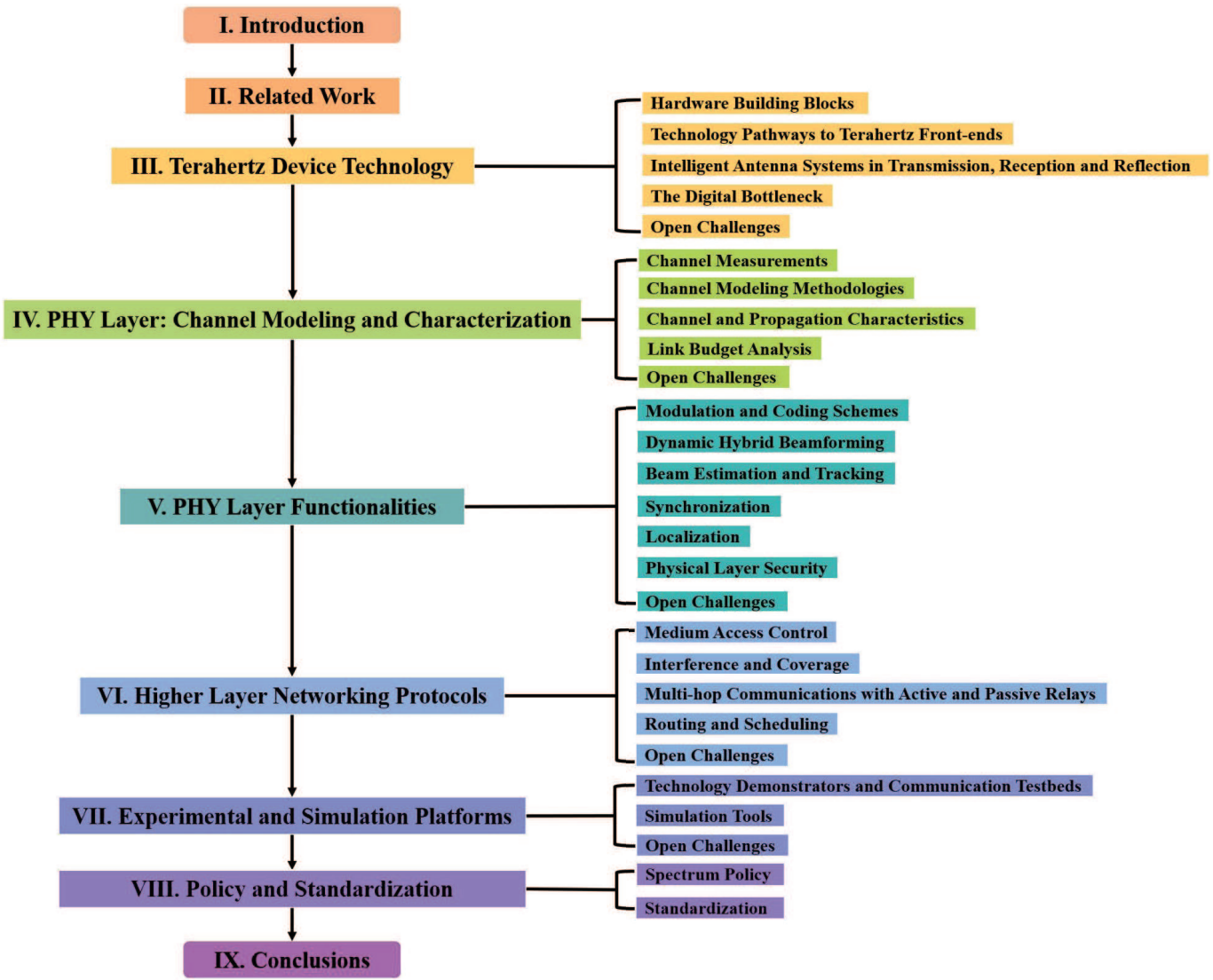}
	\caption{The structure of this paper.}
	\label{fig:flow}
\end{figure*}

Throughout the paper, we highlight, discuss, and address the grand challenge in THz communications: the limited communication distance resulting from a combination of the physics of the wireless channel and the physics of THz devices. 
On the one hand, very high free-space propagation loss, diffusely scattering, vulnerability to line-of-sight (LoS) blockage, and molecular absorption loss impose channel degradation of the THz spectrum. 
On the other hand, limited transmit power and reduced receiver sensitivity downgrade the signal strength. 
The THz device technology is investigated in Sec.~\ref{sec_device}, and the THz channel is analyzed in Sec.~\ref{sec_channel}, which will highlight the distance problem.
In our prior work~\cite{akyildiz2018combating}, four directions to tackle the crucial problem of distance limitation are investigated, including distance-aware physical layer design, ultra-massive multiple-input-multiple-output (UM-MIMO) communication, reflectarrays, and intelligent surfaces. In this work,
we will present the solutions to overcome this problem, ranging from physical-layer functionalities in Sec.~\ref{sec_PHY}, higher-layer networking protocols in Sec.~\ref{sec_network}, to experimental and simulation testbeds in Sec.~\ref{sec_testbeds}.
{Spectrum policy and standardization are described in Sec.~\ref{sec_standards}}.
The organization of this work is illustrated in Fig.~\ref{fig:flow}. 

\section{Related Work}

In recent years, the publications on THz band reached a peak where not only technical papers are being published but also many tutorials and review papers are appearing in the literature.
In this section, we give a global view of all these publications in Table~\ref{tab:related}.
Note that if any important paper is missing, it is not intentional and we apologize ahead of time.

Different research directions are highlighted in these works, including five main aspects, namely, hardware for THz communications\response{~\cite{federici2010terahertz, relatedAdd1, song2011present,teranets,Akyildiz,nagatsuma2016advances,mehdi2017thz,sengupta2018terahertz,chen2019survey,Elayan2020THz,relatedAdd2,ahi2020survey,relatedAdd4,sarieddeen2021overview}}, THz channels\response{~\cite{piesiewicz2007short,federici2010terahertz,song2011present,teranets,Akyildiz,han2018propagation,chen2019survey,rappaport2019wireless,Elayan2020THz,relatedAdd2,relatedAdd3,lemic2021survey,sarieddeen2021overview,han2021terahertz}}, THz communication techniques\response{~\cite{teranets,Akyildiz,headland2018tutorial,chen2019survey,rappaport2019wireless,Elayan2020THz,relatedAdd3,relatedAdd4,interferenceCoverageProblem,lemic2021survey,sarieddeen2021overview}}, THz networking\response{~\cite{teranets,Akyildiz,relatedAdd3,interferenceCoverageProblem,lemic2021survey}}, and THz experimental and simulation testbeds\response{~\cite{teranets,Akyildiz,chen2019survey,lemic2021survey,sarieddeen2021overview}}.

Furthermore, the number of articles surveying THz communications raised remarkably after 2017, suggesting the ceaseless increase of research attention to this untapped potential band recently. A holistic review of the THz communications that covers all the classical and novel research directions is hence motivated. In contrast to most of the existing related works, a thorough inspection of THz wireless communications is presented in this paper, containing THz devices, channels, communication techniques, networking, and testbed.
Although all these aspects were covered in~\cite{teranets,Akyildiz}, relevant work was scarce before 2014, especially about THz networking and testbeds. 
However, owing to the fast development of THz communications, more and more creative and influential solutions arise.
In this article, we revisit all the most recent research publications about THz communications, based on which the innovative open issues and potential future research directions are proposed.


\begin{table*}
	\centering
		\caption{Related tutorials, surveys and magazines.}
	\label{tab:related}
	\resizebox{\textwidth}{!}{
	\begin{tabular}{@{}p{7cm}<{\centering}ccccp{2cm}<{\centering}cc@{}}
\toprule
                                                                                                                                                       &                                 & \multicolumn{5}{c}{\textbf{Component}}                                              \\ \cmidrule(l){3-7} 
\multirow{-2}{*}{\textbf{Title and Reference}}                                                                                                         & \multirow{-2}{*}{\textbf{Year}} & Device       & Channel      & PHY Layer Functionality & Network      & Testbed      \\ \midrule
Short-Range Ultra-Broadband Terahertz Communications: Concepts and Perspectives \cite{piesiewicz2007short}                                             & 2007                            &              & $\checkmark$ &                         &              &              \\
\rowcolor[HTML]{C0C0C0} 
Review of Terahertz and Subterahertz Wireless Communications  \cite{federici2010terahertz}                                                             & 2010                            & $\checkmark$ & $\checkmark$ &                         &              &              \\
Terahertz Terabit Wireless Communication  \cite{relatedAdd1}                                                                                           & 2011                            & $\checkmark$ &              &                         &              &              \\
\rowcolor[HTML]{C0C0C0} 
Present and Future of Terahertz Communications  \cite{song2011present}                                                                                 & 2011                            & $\checkmark$ & $\checkmark$ &                         &              &              \\
TeraNets: Ultra-broadband Communication Networks in the Terahertz Band  \cite{teranets}                                                                & 2014                            & $\checkmark$ & $\checkmark$ & $\checkmark$            & $\checkmark$ & $\checkmark$ \\
\rowcolor[HTML]{C0C0C0} 
Terahertz Band: Next Frontier for Wireless Communications  \cite{Akyildiz}                                                                             & 2014                            & $\checkmark$ & $\checkmark$ & $\checkmark$            & $\checkmark$ & $\checkmark$ \\
Advances in Terahertz Communications Accelerated by Photonics  \cite{nagatsuma2016advances}                                                            & 2016                            & $\checkmark$ &              &                         &              &              \\
\rowcolor[HTML]{C0C0C0} 
THz Diode Technology: Status, Prospects, and Applications  \cite{mehdi2017thz}                                                                         & 2017                            & $\checkmark$ &              &                         &              &              \\
Propagation Modeling for Wireless Communications in the Terahertz Band   \cite{han2018propagation}                                                     & 2018                            &              & $\checkmark$ &                         &              &              \\
\rowcolor[HTML]{C0C0C0} 
Tutorial: Terahertz Beamforming, from Concepts to Realizations \cite{headland2018tutorial}                                                             & 2018                            &              &              & $\checkmark$            &              &              \\
Terahertz Integrated Electronic and Hybrid Electronic–Photonic Systems   \cite{sengupta2018terahertz}                                                  & 2018                            & $\checkmark$ &              &                         &              &              \\
\rowcolor[HTML]{C0C0C0} 
A Survey on Terahertz Communications \cite{chen2019survey}                                                                                             & 2019                            & $\checkmark$ & $\checkmark$ & $\checkmark$            &              & $\checkmark$ \\
Wireless Communications and Applications above 100 GHz: Opportunities and Challenges for 6G and Beyond    \cite{rappaport2019wireless}                 & 2019                            &              & $\checkmark$ & $\checkmark$            &              &              \\
\rowcolor[HTML]{C0C0C0} 
Terahertz Band: The Last Piece of RF Spectrum Puzzle for Communication Systems   \cite{Elayan2020THz}                                                  & 2019                            & $\checkmark$ & $\checkmark$ & $\checkmark$            &              &              \\
Wave Propagation and Channel Modeling in Chip-Scale Wireless Communications: A Survey from Millimeter-Wave to Terahertz and Optics  \cite{relatedAdd2} & 2019                            & $\checkmark$ & $\checkmark$ &                         &              &              \\
\rowcolor[HTML]{C0C0C0} 
Survey of Terahertz Photonics and Biophotonics \cite{ahi2020survey}                                                                                    & 2020                            & $\checkmark$ &              &                         &              &              \\
A Holistic Investigation of Terahertz Propagation and Channel Modeling toward Vertical Heterogeneous Networks    \cite{relatedAdd3}                    & 2020                            &              & $\checkmark$ & $\checkmark$            & $\checkmark$ &              \\
\rowcolor[HTML]{C0C0C0} 
Beyond 100 Gb/s Optoelectronic Terahertz Communications: Key Technologies and Directions  \cite{relatedAdd4}                                           & 2020                            & $\checkmark$ &              & $\checkmark$            &              &              \\
Toward End-to-End, Full-Stack 6G Terahertz Networks  \cite{interferenceCoverageProblem}                                                                & 2020                            &              &              & $\checkmark$            & $\checkmark$ &              \\
\rowcolor[HTML]{C0C0C0} 
Survey on Terahertz Nanocommunication and Networking: A Top-Down Perspective \cite{lemic2021survey}                                                    & 2021                            &              & $\checkmark$ & $\checkmark$            & $\checkmark$ & $\checkmark$ \\
An Overview of Signal Processing Techniques for Terahertz Communications \cite{sarieddeen2021overview}                                                 & 2021                            & $\checkmark$ & $\checkmark$ & $\checkmark$            &              & $\checkmark$ \\
\rowcolor[HTML]{C0C0C0} 
Terahertz Wireless Channels: A Holistic Survey on Measurement, Modeling, and Analysis \cite{han2021terahertz}                                          & 2021                            &              & $\checkmark$ &                         &              &              \\ \bottomrule
\end{tabular}
}
\end{table*}

\section{{Terahertz Device Technologies}}
\label{sec_device}
Traditionally, the lack of compact energy-efficient high-power THz transmitters and low-noise high-sensitivity receivers has limited the practical use of the THz band for communication systems. Fortunately, the so-called THz technology gap has been significantly closed in the last decade thanks to major advancement in semiconductor technologies and the development of new materials.

In this section, we first describe the key hardware blocks of a THz wireless system, namely, analog front-ends, antenna systems and digital back-ends, and list the key performance metrics. Then, we review the state of the art and open challenges for each building block. 

\subsection{Hardware Building Blocks}
The hardware building blocks of a THz wireless communication and sensing system include the analog front-ends, the antenna systems, and the digital back-ends:
\begin{itemize}
    \item The \textbf{analog front-end} at the transmitter is in charge of generating a THz carrier signal or pulse-based waveform, modulating the signal according to the information to be transmitted, amplifying the signal power, and filtering out-of-band emissions (e.g., harmonics) prior to radiation. Reciprocally, the analog front-end at the receiver is in charge of detecting a THz signal, filtering the out-of-band noise, amplifying the useful signal and recovering the information. The key performance metrics of a front-end include the frequency bands of operation, modulation bandwidth, transmission power, receiver sensitivity, and amplitude and phase noise.
    
    \item The \textbf{antenna system} is in charge of converting on-chip modulated THz signals at the transmitter front-end into free-space propagating EM signals and, reciprocally, at the receiver, coupling free-space radiation onto the on-chip front-ends. The antenna system can be integrated by one or more antennas, in combination with lenses or reflecting structures. The key performance metrics of an antenna system includes radiation efficiency, directivity gain and beamdwidth. 
    
    \item The \textbf{digital back-end} is in charge of generating the control signals for the analog front-ends, digitally processing the information signals to be transmitted and received through the analog front-ends and, ultimately, serving as an interface between digital computing devices and the analog wireless channel. The main elements of the digital back-end are the digital to analog converters (DACs) at the transmitter and analog to digital converters {(ADCs)} at the receiver. Their key performance metrics include the sampling frequency and the sampling resolution. 
\end{itemize}

To unleash the potential of the THz spectrum, a THz communication system should be able to operate at one or more frequency windows within the THz band, with modulation bandwidths in excess of 10~GHz at very least (i.e., two orders of magnitude more than in current 5G systems), a combination of transmission power, antenna gains, and receiver sensitivity that meets the link budget requirements according to the propagation conditions and the targeted communication distance (Sec.~\ref{sec:link_budget}), and with low amplitude and phase noise powers. High power efficiency, low energy consumption and reconfigurability are also critical to the development of intelligent, dynamic and sustainable THz communication and sensing systems for 6G and Beyond.

\subsection{Technology Pathways to Terahertz Front-ends}

\begin{figure}[!h]
	\centering
	\includegraphics[width=\linewidth]{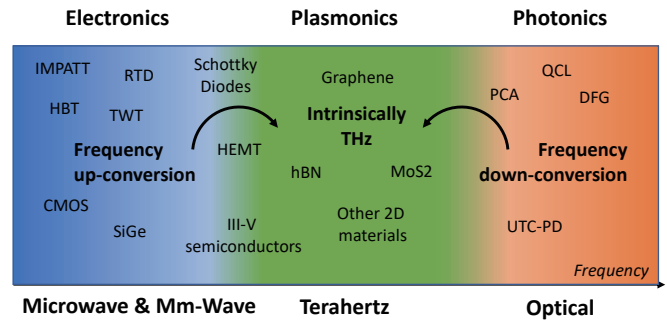}
	\caption{Technology pathways to THz front-ends.}
	\label{fig:technologies}
\end{figure}

There are currently three main technology pathways or approaches to the development of THz systems, namely, the electronic, the photonic and the plasmonic approach (Fig.~\ref{fig:technologies}). Next, we summarize the state of the art for each approach. 

\subsubsection{Electronic Approach}
The goal of this approach is to push the limits of existing technologies for microwave and mm-wave devices to enable THz systems. Resonant Tunneling Diodes (RTDs)~\cite{diebold2016modeling,kasagi2019large}, IMPact ionization Avalanche Transit Time (IMPATT) diodes~\cite{biswas20181,banerjee2020thz}, and Traveling Wave Tubes (TWTs)~\cite{baig2017performance,hu2019demonstration} are examples of the electronic approach. {In all these cases, the generated signal frequency is of a few hundreds of GHz. In terms of power, sub-mW emissions are expected from RTDs and IMPATT diodes, where as up to a few Ws have been demonstrated with bulkier TWTs.} Besides all these, the most common technique to generate a THz carrier signal in an electronic approach is the use of \textbf{frequency multiplying chains to up-convert a lower-frequency signal} (generally from the microwave and millimeter-wave band). Frequency multiplying chains are integrated by concatenated frequency doublers and/or triplers and are generally followed by a broadband mixer in a heterodyne architecture. The same architecture can be used reciprocally at the receiver to down-convert the received signal back to an intermediate frequency (IF) or directly to baseband, where the digital block enters the game (Sec.~\ref{subsec:digital}). Frequency multipliers can be built utilizing different device technologies:
\begin{itemize}
    \item \textbf{Silicon CMOS and Silicon-Germanium (SiGe) BiCMOS} technologies offer high level of integration with existing electronic systems. However, their maximum transmission power is up to \emph{a few milliwatts per element at frequencies of up to a few hundreds of GHz}~\cite{hara2018300,kenneth2019opening}.
    \item To increase both the transmission power and the maximum frequency of operation, \textbf{III-V semiconductors,} such as Indium Phosphide (InP) and Gallium Arsenide (GaAs), are being utilized to fabricate high-electron mobility transistors (HEMT), heterojunction bipolar transistors (HBT) and Schottky diodes to be utilized in frequency multiplying chains. As of today, transmission power of \emph{hundreds of milliwatts at a few hundreds GHz and up to a few mWs above 1~THz} have been experimentally demonstrated~\cite{leong2017850,siles2018new}.
\end{itemize}
Independently of the specific fabrication technology, the frequency multiplication and mixing process is \emph{highly non-linear}. This introduces several challenges for THz communications, including high waveform distortion, very high sensitivity to peak to average power ratio (PAPR), and very high phase noise. Moreover, if utilizing transistor-based multipliers as opposed to diode-based multipliers, high gain is generally achieved at the cost of reduced bandwidth. For example, the bandwidth of the InP HEMT-based frequency-multiplied transceivers is generally less than 10~GHz~\cite{leong2017850}, as opposed to the several tens of GHz of Schottky-diode-based systems~\cite{siles2018new}. 

To both increase the transmission power as well as to enable mobile THz networking, \emph{electronic arrays} are being intensively explored. For example, transceiver arrays with 8 parallel channels at 140~GHz have been experimentally demonstrated~\cite{simsek2020140ghz,abu2021end}. Each channel can be utilized to drive independent on-chip antennas in a fully digital MIMO approach or in conjunction with analog antenna arrays to implement hybrid beamforming architectures (Sec.~\ref{DAoSA}). While on-chip THz antennas are very small thanks to the very small wavelength of THz signals (Sec.~\ref{subsec:antennas}), the challenge is to integrate all the electronic components in a thermally stable package~\cite{song2017packages,alonso2019micromachining}.


As of 2022, we expect the first commercial THz systems to operate in the lower end of the G-band (from 110~GHz to 300~GHz), and given current technology maturity, these are likely to be based on frequency multiplying chains.

\subsubsection{Photonic Approach}
The goal of this approach is to push the limits of photonic techniques utilized in optical wireless systems (OWC) down to the THz band. Currently, there are mostly three approaches to the photonic generation, modulation and detection of THz signals:
\begin{itemize}
    \item \textbf{Quantum Cascade Lasers (QCLs):} These are semiconductor lasers that emit low-energy photons corresponding to far-infrared and THz frequencies~\cite{williams2007terahertz,wang2016high}. 
    The frequency range of operation of QCLs is generally above 2-3~THz. While output powers of up to a few hundreds of mWs have been demonstrated at cryogenic temperatures, the open challenge is to ensure practical power figures when operating at room temperature~\cite{lu2019room,khalatpour2021high}. This currently limits their utilization in practical applications. 
    \item \textbf{Frequency-difference Generation (DFG):} The fundamental idea of DFG systems is the reciprocal to that of frequency-multiplied electronic systems: a THz carrier signal can be generated by photomixing two optical carrier signals whose wavelength differs by an increment that corresponds to the target THz signal~\cite{nagatsuma2016advances,sung2021design}. These systems usually consist of two narrow-band lasers followed by a uni-travelling photodiode (UTC) which acts as a down-converter to THz frequencies~\cite{renaud2017antenna,relatedAdd4}. For the time being, these systems have been demonstrated to perform at frequencies of a few hundreds of GHz, with powers of less than 10~mW.
    \item \textbf{Photoconductive antennas (PCAs):} Another technique to downconvert optical signals to the THz band relies on the use of photoconductive antennas~\cite{burford2017review,yardimci2018nanostructure}. A photoconductive antenna generally consists of a conventional THz metallic antenna printed on top of a photoconductive substrate. When illuminated by an optical signal (normally an optical pulse), photocarriers are excited at the gap of the antenna. A DC bias field is then used to accelerate the carriers along the antenna structure, resulting in the emission of THz photons. The main challenge of this technique relates to the low conversion efficiency and the resulting low emitted power (less than a milliwatt at hundreds of GHz).
\end{itemize}

The transmission power of photonic THz systems is at least one order of magnitude lower than that of electronic systems and, while the development of photonic THz arrays is an active field of research, these have not been experimentally demonstrated. Nonetheless, the signals generated through frequency down-conversion exhibit higher spectral purity and lower amplitude and phase noise. Moreover, existing large-bandwidth optical modulators can be leveraged to generate high bandwidth THz signals. In light of these, a potential application in which photonic THz systems are likely to be found in the future is in wired fiber optical to wireless THz backhaul applications. 
{The feasibility of such blending of wired fiber optical to wireless THz links has been discussed in a  in-depth literature survey~\cite{Boulogeorgo2018THz}, by taking into account transceiver radio-frequency (RF) front-end devices, channel modeling schemes, PHY layer functionalities, medium access control (MAC) protocols, as well as signal processing designs.}

{Photonic transmitters are generally used in conjunction with electronic receivers generally based on frequency-multiplying chains in heterodyne configurations as discussed in the previous section. This is due to the lack of an optical device that can up-convert information-bearing THz signals to the optical spectrum, where to be processed by leveraging existing optical technologies. Recently, the use of a DFG system to generate an on-chip LO to down-convert the received THz signals to an electronic IF signal has been shown in~\cite{harter2019wireless}, introducing a new concept of how photonic technologies could be leveraged at the receiver.}


\begin{figure*}[t]
	\centering
	\includegraphics[width=\linewidth]{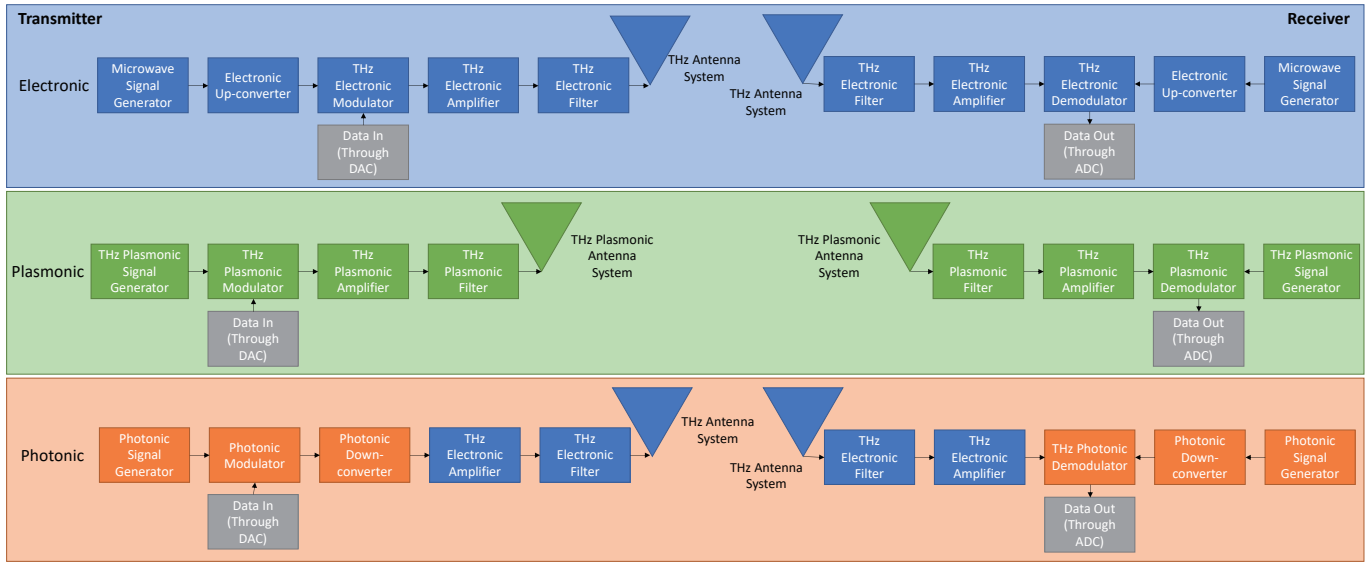}
	\caption{{Block diagram for the three most common transceiver architectures based on: (top) Electronic; (middle) Plasmonic; and (bottom) Photonic.}}
	\label{fig:architectures}
\end{figure*}

\subsubsection{Plasmonic Approach}

The goal of this approach is to create devices that intrinsically operate at THz frequencies, i.e., without the need to up-convert from the microwave range or down-convert from optics, by leveraging the properties of plasma waves. While some of the first concepts relating to plasmonic generation of THz signals date back over 20 years~\cite{dyakonov1993shallow,knap2004plasma}, the discovery and utilization of \textbf{new two dimensional nanomaterials, such as graphene}~\cite{novoselov2012roadmap}, has opened the door to THz nano-devices able to operate at room temperature. There are several properties of graphene that makes them especially attractive for THz communications. On the one hand, the very high electron mobility
at room temperature enables the propagation of \textbf{surface plasmon polariton (SPP) waves,} or global oscillation of electrical charges at the interface of graphene and a dielectric material. By leveraging the properties of SPP waves, new devices able to generate~\cite{jornet2014graphene,crabb2021hydrodynamic}, radiate~\cite{jornet2010graphene,tamagnone2012analysis,jornet2013graphene,abadal2015time} and detect~\cite{vicarelli2012graphene,bandurin2018resonant} THz signals have been developed~\cite{jornet2014graphene,jornet2013graphene}. 
On the other hand, graphene is a highly tunable material, a property that is leveraged to design ultrabroadband modulators in our previous works~\cite{singh2016graphene,crabb2021chip}. This is necessary to ultimately realize the 1~Tbps data-rates envisioned for 6G systems. Moreover, to enhance some of these properties, graphene can be combined with III-V semiconductors or with other 2D materials, such as hexagonal boron nitride (hBN) or molybdenum disulfide (MoS2).

Plasmonic THz devices \emph{naturally operate at frequencies above hundreds of GHz and few THz with bandwidths in excess of 10\% of their carrier frequency}. In addition, they are very small, i.e., in the order of hundreds of nanometers or a few micrometers. Note that this is much smaller than the THz wavelength, because of the {plasmonic confinement factor.} The very small size plays a dual role in THz communication systems. On the one hand, \emph{even if energy efficient, the total radiated power by a single device is very low, in the orders of a few microwatts at most.} On the other hand, their very small size allows their adoption the nanoscale applications discussed in the introduction~\cite{akyildiz2010electromagnetic,nanothings}, but also their integration in very dense compact arrays~\cite{akyildiz2016realizing, singh2020design}. 
{More specifically, THz waves can be radiated via both nano-dipole and nano-patch antennas with the antenna size of several hundreds of nanometers~\cite{jornet2010graphene}.
In addition, with a very small electrical size, graphene is promising to outperform the metal counterparts in THz radiation efficiency~\cite{tamagnone2012analysis}.
Compared to the metallic antennas of the same size, graphene-based nano-antennas can operate at much lower frequencies like the THz band as opposed to optical frequencies~\cite{jornet2013graphene}.
With the comparable performance of metallic antennas, the tradeoff between pulse dispersion and radiation efficiency of graphene-based antennas is further discussed in~\cite{abadal2015time}.
}

Nevertheless, compared to the electronic and photonic approaches, which have been refined over decades of research, the plasmonic approach is still in its early stages. {Mostly, while graphene had been theoretically explored since the 19th century, it was not until 2004 when it was experimentally obtained and characterized~\cite{geim2007rise}. This is a material than then needs to be transformed into actual devices (e.g., graphene-based transistors). We have been granted patents on graphene-based plasmonic transceivers~\cite{jornet2014graphene}, antennas~\cite{jornet2013graphene}, and arrays~\cite{akyildiz2016realizing} for THz band communications, which we believe that the fabrication will help the community to realize productions of devices. Comparatively, silicon-based transistors were developed in the 1950s. While there has been tremendous progress since 2004, key challenges for the graphene-based plasmonic approach are still present. Above all, critical graphene properties such as the high electron mobility and long electron relaxation time have mostly been observed when graphene is isolation, but quickly drop when in contact with other conventional materials. This is what is motivating the development of heterogeneous structures that combine different 2D and 3D materials, but this research is still mostly at the material or single device levels.} In our vision, the plasmonic approach has the potential to enable compact, energy-efficient and intelligent distributed THz systems.




Other non-conventional devices for THz communications include \textbf{phase-change materials,} including vanadium dioxide ($\mathrm{VO_2}$) and liquid crystals~\cite{wang2018liquid,jeong2020dynamic}. Specifically, phase-change materials can be utilized as tunable elements at the transmitter and the receiver to support high-speed modulation (in addition to phase controllers in antenna systems, discussed in Sec.~\ref{subsec:antennas}). Furthermore, the application of \textbf{superconductors} like $\mathrm{Bi_2Sr_2CaCu_2O_{8+\delta}}$ (BSCCO) also enable tunable plasmonic THz sources~\cite{delfanazari2020integrated}.

Overall, the plasmonic approach to THz communications is a less mature and conservative path than the electronic and photonic approaches, with \emph{high risk but also very high rewards,} arising from true THz operation, ultrabroadband bandwidths, and support for high reconfigurability and tunability.

As a summary of this section, the block diagrams for the three most common analog front-end archictectures for THz wireless systems is shown in Fig.~\ref{fig:architectures}. 



\subsection{Intelligent Antenna Systems in Transmission, Reception and Reflection}
\label{subsec:antennas}
Antennas are needed to convert the on-chip signals into wirelessly propagating waves. In contrast to analog front-ends, where there are clear divisions between electronic, photonic and plasmonic systems, antenna systems are much more flexible. For example, conventional antennas are generally used in conjunction with electronic and photonic front-ends, whereas plasmonic antennas are needed for plasmonic front-ends. Similarly, optical lenses can be used in conjunction with any type of antenna. We explain these in detail next.

The very small wavelength of THz signals enables the development of very small conventional THz antennas. For example, a 150~$\mu$m long printed dipole or patch antenna resonates at approximately 1~THz and exhibits a quasi-omnidirectional radiation pattern. Moreover, if adopting graphene and/or other plasmonic materials, the plasmonic confinement factor can be leveraged to design THz antennas which are even smaller. For example, a graphene-based plasmonic nano-patch antenna designed in our previous work to resonate at the same 1~THz would be just 1~$\mu$m long and 10~nm wide, i.e., two orders of magnitude smaller than the metallic antenna~\cite{jornet2013graphene}. The very small size of these antennas is what enables their application in wireless networks on chip or even wearable and implantable devices (Fig.~\ref{fig:applications}).

Nevertheless, this very small size also results in a very small antenna effective area and, correspondingly, very high spreading losses (Sec.~\ref{sec:channel char}). Increasing the effective area of THz antennas automatically leads to antennas that no longer radiate omnidirectionally, but are highly directional. For the time being, there are mostly two main types of directional antenna systems:
\begin{itemize}
    \item \textbf{Fixed directional antennas:} These include horn antennas, with typical directivity gains in the order of 20-25~dBi (or gain in dB over an isotropic antenna); Cassegrain dish reflector antennas, with directivity gains around 40-55~dBi; as well as more sophisticated designs such as multi-reflector antennas~\cite{fan2017development}, with gains of up to 30-35~dBi. Many of these are commercially available~\cite{commercialAntenna1,commercialAntenna2}. Other less conventional fixed directional radiating structures include leaky wave antennas, whose direction of radiation depends on the frequency of excitation, which opens the door to new ways of spatial and frequency multiplexing~\cite{esquius2014sinusoidally,karl2015frequency}.
    \item \textbf{Beamforming antenna arrays:} These can be either fixed or dynamic, and can be built either with traditional metals or by leveraging new nanomaterials. For example, on the one hand, metallic antenna arrays utilized in conjunction with electronic front-ends have been demonstrated at frequencies under 300~GHz~\cite{sengupta20120,tang201465nm, simsek2020140ghz} with up to 16 controllable elements. On the other hand, in conjunction with plasmonic front-ends, very large graphene-based plasmonic antenna arrays with up to 1,024 elements have been proposed~\cite{akyildiz2016realizing,shen2019planar,singh2020design}. The latter are at the basis of ultra-massive MIMO systems, as we discuss in Sec.~\ref{DAoSA}.
\end{itemize}

In both cases, \textbf{dielectric lenses} can be utilized to further increase the gain and dynamic capabilities of directional THz antenna systems. For example, a silicon lens printed on top of an on-chip quasi-omnidirectional antenna can provide up to 30~dBi of gain~\cite{llombart2011novel}. Similarly, commercially-available lens-integrated horn antennas can achieve gains in the range of 40-55~dBi, in a more compact form factor than that of dish reflector antennas. Moreover, lenses can also be utilized in conjunction with arrays. For example, in~\cite{yurduseven2016dual}, an array of lenses deposited on top of an array of of on-chip antennas is experimentally demonstrated. Alternatively, a single large footprint lens can be utilized on top of a beamforming antenna array, for which the beamforming weights would need to be designed accounting for the lens-corrected propagation path.

In addition to fixed lenses, \textbf{programmable lenses based on metamaterials} can also be utilized at THz frequencies~\cite{tao2009reconfigurable,tao2010recent,ju2011graphene,pacheco2017experimental}. Metamaterials and metasurfaces are engineered structures (in 3D and 2D, respectively) designed to exhibit EM properties not commonly found in nature. Metamaterials are composed by well-defined arrangements of meta-atoms, which are generally metallic structures. By definition, the size of the meta-atoms is much smaller than the wavelength corresponding to the frequency at which the structure is designed to operate. This is a key difference from traditional metallic antenna arrays, with which many times metasurfaces are compared. Perhaps more importantly, antennas are transducers that convert electrical energy into EM energy and vice-versa, as opposed to metamaterials, in which the input and the output are EM signals. Besides lensing, transmit metasurfaces can also be utilized to perform communication functions of the chip, such as modulation~\cite{sensale2012broadband}, polarization manipulation~\cite{cheng2014ultrabroadband}, and holographic projections~\cite{venkatesh2020high}.

Besides transmission and reception, antenna arrays and metasurfaces can also be utilized in reflection and, thus, as enablers of \textbf{reflecting intelligent surfaces (RISs).} RISs are a key technology to enable intelligent propagation environments in 6G systems~\cite{nie2019intelligent,akyildiz20206g}. While RISs at lower frequencies are mostly used to increase the achievable data-rates, at THz frequencies RISs are a critical infrastructure to overcome the need for line-of-sight (LoS) propagation (Sec.~\ref{sec:channel char}). RISs can precisely control the reflection of incoming THz signals in specular and non-specular paths, while changing the polarization or adding an additional level of modulation if needed. Fixed THz reflecting surfaces based on reflecting antenna arrays or reflect-arrays as well as metasurfaces have been experimentally demonstrated for years~\cite{he2013generation,dong2015terahertz,gao2015broadband}. The challenge is to make them reconfigurable. In contrast to lower frequency bands, where tunable capacitors or diodes can be utilized to change the reflection coefficient of individual antenna elements or the behavior of individual meta-atoms, at THz frequencies new tunable elements are needed. The tunability of graphene can be leverage for this purpose. For example, in~\cite{carrasco2013reflectarray}, a graphene-based plasmonic reflect-array was proposed. However, graphene is not a good reflecting material compared to traditional metals. To overcome this channel, we have proposed hybrid graphene-metal structures recently ~\cite{singh2020designr1,singh2020hybrid}, which combine the tunability of graphene with the reflectivity of gold. On the metamaterial front, we have proposed hypersurfaces~\cite{liaskos2018new,liaskos2018using} to enable reconfigurable metasurfaces that can achieve different functionalities across the spectrum, including the THz band~\cite{dash2021switched}. {A summary of the key radiating structures and their properties is provided in Table~\ref{table:antennas}, including usage in transmission (Tx), reception (Rx) and reflection (Rf); size; and operating modes, such as beamforming (BF), spatial multiplexing (SM), absorption (AB), polarization manipulation (PM), and wavefront engineering (WE).}

\begin{table*}
    \centering
    \caption{Summary of the properties of different antenna systems.}
\begin{tabular}{|c|ccc|c|ccccc|}
\hline
\multirow{2}{*}{\textbf{}}                 & \multicolumn{3}{c|}{Usage}                                         & \multirow{2}{*}{Size}      & \multicolumn{5}{c|}{Operating mode}                                                                                            \\ \cline{2-4} \cline{6-10} 
                                           & \multicolumn{1}{c|}{Tx}     & \multicolumn{1}{c|}{Rx}     & Rf     &                            & \multicolumn{1}{c|}{BF}     & \multicolumn{1}{c|}{SM}     & \multicolumn{1}{c|}{AB}     & \multicolumn{1}{c|}{PM}     & WE     \\ \hline
\textbf{Antenna and antenna arrays}        & \multicolumn{1}{c|}{\cmark} & \multicolumn{1}{c|}{\cmark} & \cmark & Element $\simeq \lambda/2$ & \multicolumn{1}{c|}{\cmark} & \multicolumn{1}{c|}{\cmark} & \multicolumn{1}{c|}{\xmark} & \multicolumn{1}{c|}{\xmark} & \cmark \\ \hline
\textbf{Dielectric lenses and lens arrays} & \multicolumn{1}{c|}{\cmark} & \multicolumn{1}{c|}{\cmark} & \cmark & Element $\geq \lambda$     & \multicolumn{1}{c|}{\cmark} & \multicolumn{1}{c|}{\cmark} & \multicolumn{1}{c|}{\xmark} & \multicolumn{1}{c|}{\cmark} & \cmark \\ \hline
\textbf{Metasurfaces}                      & \multicolumn{1}{c|}{\cmark} & \multicolumn{1}{c|}{\cmark} & \cmark & Element $\ll \lambda/2$    & \multicolumn{1}{c|}{\cmark} & \multicolumn{1}{c|}{\cmark} & \multicolumn{1}{c|}{\cmark} & \multicolumn{1}{c|}{\cmark} & \cmark \\ \hline
\end{tabular}
\label{table:antennas}
\end{table*}

In addition to the spatial properties of these radiating structures (e.g., number of beams, beamwidth, etc.), \textbf{the antenna bandwidth plays a key role.} With only few exceptions (such as in impulse-radio ultra-wide-band systems), antennas utilized in commercial communication systems are generally narrowband, i.e., the bandwidth of their frequency response is much less than 10\% of the carrier signal. Antenna systems effectively acts as filters and, thus, can introduce distortion to the transmitted waveforms if these exceed the antenna bandwidth. Such dispersive effect, many times attributed to the channel, is in fact an antenna factor and, thus, the design of ultrabroadband antennas becomes critical. Even if the radiating elements themselves exhibit large bandwidth, the other building blocks of antenna or meta-atom arrays need to be large bandwidth. For example, in antenna arrays, broadband phase controllers or, reciprocally, time-delay lines are needed to prevent beam squint~\cite{han2021hybrid}.

Finally, additional opportunities for THz antenna systems include the development of structures that can enable the \textbf{engineering of wavefronts} different than the traditional Gaussian beams. For example, self-healing Bessel beams~\cite{liu2016anomalous} or bending Airy beams~\cite{liu20163d} can provide additional capabilities to intelligent THz propagation environments. Moreover, on top of the wavefronts, orbital angular momentum (OAM) can be defined~\cite{zhu2014experimental,su2020multipath}. Often compared to MIMO, OAM creates intrinsically orthogonal parallel spatial channels, without the need of sophisticated digital signal processing and working even (or especially) in LoS propagation conditions. Such OAM modes can be utilized to increase the capacity of THz systems without increasing the system bandwidth.


\subsection{The Digital Bottleneck}
\label{subsec:digital}

Enabling 1~Tbps links requires not only the development of analog front-ends and antenna systems that can operate at higher frequencies and with larger bandwidths, but also the development of a digital signal processing back-end that can generate at the transmitter and process at the receiver the information at such very high date. For the time being, all {digital signal processing (DSP)} solutions rely on electronic devices, mostly because of all the commercial computing cores are electronic systems. While there are early concepts of fully optical processors~\cite{xie2018programmable,tang2021ten} and even plasmonic processors~\cite{fu2012all,gogoi2016all}, these are not likely to become a reality within 6G systems. Therefore, we focus on electronic DSP systems.

Common with the fastest wired communication systems over optical fiber, the main bottleneck for ultra-broadband digital THz communications is posed by \textbf{the performance of DACs and ADCs.} More specifically,
\begin{itemize}
    \item At the transmitter, state-of-the-art DACs able to sample at frequencies in excess of 100~Giga-samples-per-second (GSaps) have been experimentally demonstrated. For example, in~\cite{nagatani2018256}, two 128~GSaps DACs are multiplexed to effectively sample at 256~GSaps. This translates to a signal bandwidth of at most 128~GHz (and practically much less if meaningful oversampling gain is needed). However, this comes at the cost of low sampling resolution (only 2-bits) very high power requirements, thermal dissipation and large footprint.
    \item The situation is not that different at the receiver. Generally, higher sampling frequency and higher resolution ADCs have been developed, many times driven by the scientific sensing community performing remote sensing experiments at THz frequencies. Time-interleaving is also the solution commonly adopted to increase the sampling frequency of ADCs~\cite{buchwald2016high}. Again, this comes at the cost of strict power, thermal and space requirements.
\end{itemize}
Other time- and frequency-multiplexed systems that can operate at 256~GSaps with 8 bits resolution are at the basis of commercial-grade equipment utilized in experimental testbeds (discussed in Sec.~\ref{sec_testbeds}).

To overcome the digital bottleneck, different strategies are being explored. On the one hand, the analog system bandwidth is divided into narrower channels in our previous work~\cite{han2015multi}, to be individually digitally processed and analogically frequency-multiplexed at the transmitter and demultiplexed at the receiver~\cite{ariyarathna2020real}. Alternatively, instead of sacrificing the sampling frequency, we have utilized a large number of very low resolution ADCs to reliably demodulated sophisticated waveforms despite heavy quantization in large arrayed systems~\cite{han2021hybrid}. Clearly, the development of faster DACs, ADCs and {DSP} algorithms will benefit not only THz communication systems, but any high speed wired or wireless communication and sensing technology.

\subsection{{Open Challenges}}
\label{subsec:device_challenges}

In addition to the individual challenges related to analog front-ends, antenna systems and digital back-ends previously highlighted in each sub-section, there are {open challenges} that affect all of the building blocks, including
\begin{itemize}
    \item {\textbf{AI-driven Smart Hardware:} In order to enable dynamic communication (Sec.~\ref{sec_PHY}) and networking (Sec.~\ref{sec_network}) solutions needed to support the applications of THz networks (Fig.~\ref{fig:applications}, Fig.~\ref{fig:nanoapp}, Fig.~\ref{fig:jcs}), agile reconfigurable hardware orchestrated by AI engines are needed. Reconfigurability has different implications for different hardware components. At the front-end level, reconfigurability implies tunability of the different analog transceiver blocks, such as the signal generators, filters, amplifiers and modulators. At the antenna system level, reconfigurability refers to the possibility of operating with different beams and wavefronts dynamically. From the digital perspective, reconfigurability involves the addition of control interfaces to each of the aforementioned blocks, which need to be operated in real-time in parallel to the fast data-converters. In our vision, the creation of agile systems will benefit from the adoption of materials and devices that are intrinsically tunable at high frequencies, such as graphene, and their hybrid integration with other electronic and photonic structures.} 
	\item \textbf{Heterogeneous integration and fabrication}.
	Different of the aforementioned building hardware blocks rely on different material and device technologies, ranging from silicon CMOS for high-density integration, to III-V semiconductors for power or graphene for tunability and bandwidth. Heterogeneous integration refers to not only the combination of different components based on different materials, but also co-design of the different blocks, exploration of impact of different blocks on each other, and definition of scalable and cost-effective fabrication processes. In our vision, to achieve optimal trade-offs among multiple metrics including frequency, bandwidth, power, cost, and energy-efficiency, the joint consideration of diverse materials and fabrication, containing photonics-based, electronics-based, nanomaterials-and-plasmonics-based approaches, is a promising research direction for practical THz communication systems.
	
	\item \textbf{Terahertz devices for extreme environments:}
	As mentioned in the introduction, the applications of THz communication networks range from nanoscale intra-body or wearable networks to inter-satellite links in space. Different environments imporse very different operation conditions. Inside the body, the use bio-compatible layers to \emph{disguise} non-biological structures from the immune system are needed. Similarly, the THz devices used in space bear noticeable differences from the ones in the terrestrial scenarios. It is relevant to note, however, that there have been satellites orbiting the Earth with THz systems (mostly power detectors) for over a decade~\cite{mehdi2017thzinstru,mehdi2017thz}. These are based on III-V Schottky diode technology. Space-proofing other of the needed technologies for extreme space applications is a necessary step to enable applications such as TeraSpace.
\end{itemize}

\section{PHY Layer: Channel Modeling and Characterization}
\label{sec_channel}

Besides practical THz devices, the realization of THz communication requires a thorough investigation of the THz channels as well. 
Unlike the low-frequency electromagnetic waves, the distinctive attributes of the THz spectrum have motivated novel techniques to model and measure the THz channels.
In particular, the high propagation loss, the frequency-dependent molecular absorption, among others, bring  challenges to the traditional channel modeling.
In this section, we explore channel measurement methods, study and assess channel modeling schemes,  analyze channels and corresponding propagation characteristics, and present open research challenges.

\subsection{Channel Measurements}
\label{channel measurement}
The first studies of THz wave propagation were largely theoretical and mostly driven by the atmospheric sensing community. Radiative transfer theory is found at the basis of closed-form analytical models for THz wave-matter interactions, which play a critical role in the propagation of THz signals. Similarly, the first complete channel models for the THz band, including our own~\cite{jornet2011channel,Han2015multiray}, are derived through the combination of physics, electromagnetics and communication theory.

Among other techniques including EM wave theory and full-wave simulation, channel measurement is essential to characterize the spectrum features and develop channel models. Recently, with the progressive narrowing of the THz technology gap, experimental measurement campaigns in the lower THz band have been conducted~\cite{stModelSVTHz,chen2021channel,ju2021millimeter,abbasi2021ultra}. Such experimental measurements play a dual role: on the one hand, they can be used to validate the analytical physics-driven models. On the other hand, empirical data-driven models, much simpler than their physical counter-parts, can be developed. The latter are generally preferred by standardization bodies, such as the 3GPP. 

\begin{figure*}[t]
\centering
        \includegraphics[width=1\textwidth]{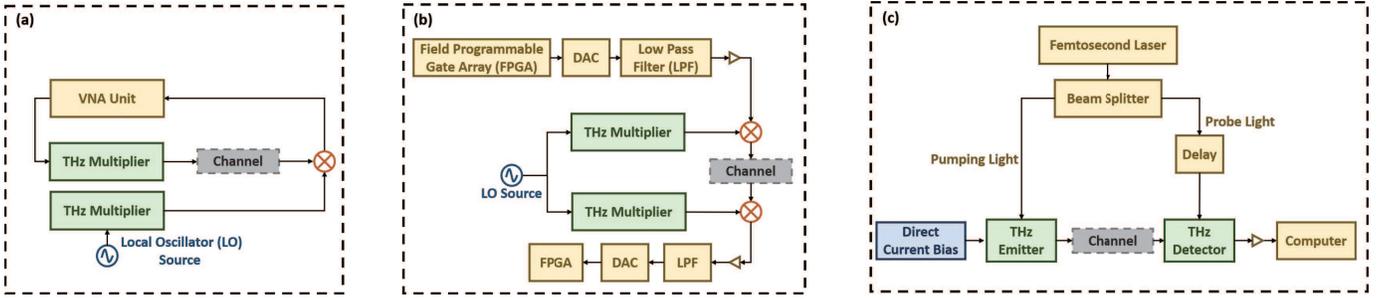} 
        \caption{Channel measurement techniques. (a)  VNA-based method, (b) sliding correlation method, and (c) TDS method.}
        \label{fig:channelmeasure}
\end{figure*}

Although multiple measurement approaches have been practically utilized for a long time, THz channel measurement systems need to measure wideband characteristics, support long-distance sounding given the high path loss, and consume reasonable time by scanning the spatial domain even with high-directivity antennas.
We describe the three main THz channel sounding techniques next (Fig.~\ref{fig:channelmeasure} {and Table~\ref{tab: channelMeasurement}}).

\begin{table*}[t]
\caption{THz channel measurement techniques~\cite{han2021terahertz}.}
\resizebox{\textwidth}{!}{
\begin{tabular}{|c|p{5cm}<{\centering}|p{5cm}<{\centering}|c|}
\hline
\textbf{} & \textbf{VNA-based}                                                                                         & \textbf{Sliding correlation}                                                                                                      & \textbf{Direct pulse}                                                                                  \\ \hline
Domain                  & Frequency                                                                                                  & Time                                                                                                                              & Time                                                                                                   \\ \hline
Signal                  & Single carrier                                                                                             & Auto-correlated sequence                                                                                                          & THz pulse                                                                                              \\ \hline
Distance                & Limited by physical link (can be compensated by RoF extension)                                             & Can support long-distance measurement                                                                                             & Limited by pulse power                                                                                 \\ \hline
Synchronization         & Implicit                                                                                                   & Required                                                                                                                          & Required                                                                                               \\ \hline
Strengths               & \begin{tabular}[c]{@{}c@{}}Large frequency range and bandwidth;\\ High time-domain resolution\end{tabular} & High measuring speed                                                                                                              & Large measuring bandwidth                                                                              \\ \hline
Weaknesses              & Low measuring speed                                                                                        & \begin{tabular}[c]{@{}p{5cm}<{\centering}@{}}Small bandwidth;\\ High complexity;\\ High requirement on synchronization and symbol rate\end{tabular} & \begin{tabular}[c]{@{}c@{}}Large size;\\ Low transmit power;\\ Limited measuring distance\end{tabular} \\ \hline
Scenarios/Applications  & Short distance                                                                                             & Long distance                                                                                                                     & Very short distance                                                                                    \\ \hline
\end{tabular}
}
\label{tab: channelMeasurement}
\end{table*}

\subsubsection{Frequency-domain Vector-Network-Analyzer-based Method}
A Vector Network Analyzer (VNA) is commonly utilized to measure the frequency response of a linear system one frequency at a time. VNAs are commonly used to characterize discrete electrical/RF components, and can be utilized to measure the wireless channel by connecting antennas at the transmitting and receiving ports. The VNA suffers from low output power, high noise, limited distance, as well as long measurement times. The first two issues, i.e., low output power and high noise, can be mitigated by exploiting external power and low-noise amplifiers. The distance can be extended via optical fiber and directional transmissions. However, the long measurement times  still hinder the application of VNA in dynamic THz channels~\cite{measureVNA}.

Based on this VNA method, we have built a 140 GHz channel measurement system with Huawei, enabling the end-to-end measurement solutions for an indoor meeting room~\cite{measureVNACamp1}. Using this system,  the channel parameters, as well as the temporal and spatial distributions of multipath scenarios, are investigated by Shanghai Jiao Tong University (SJTU)~\cite{measureVNACamp1}. Furthermore, we have improved this system to support channel measurements at 220 GHz for LoS and Non-Line-of-Sight (NLoS) paths in an office and a hallway~\cite{measureVNACamp2}. Similarly, the University of Southern California (USC) has utilized the VNA based method to create a channel measurement system for urban scenarios at approximately 140 GHz~\cite{measureVNACamp3}. To expand the transmission distance  from 8--10m to 100 m was realized through the above mentioned optical fibers in~\cite{measureVNACamp3}. Further, the measurements up to 220 GHz band is realized in~\cite{measureVNACamp4}. The VNA-based measurements at 300 GHz have been conducted at Georgia Tech  for a desktop~{\cite{measureVNACamp5}}, a computer motherboard~{\cite{measureVNACamp6}}, and a data center~{\cite{measureVNACamp7}}.

\subsubsection{Time-domain Sliding Correlation Method}
Another channel measurement approach is based on the transmission of a signal whose autocorrelation is a Dirac delta function, where the channel can be considered time-invariant. Such a correlation-based technique allows the exploitation of high transmit power with a low PAPR. The advantage of this method is that real-time measurements can be realized. However, there exist two main drawbacks with this method. On the one hand, the power spectral density tends to be distributed unevenly. On the other hand, ultra-high sampling rate DACs/ADCs are needed. The second problem can be addressed due to the fact that the sliding correlation method expands the measurement duration and accordingly increases the SNR because different chip rates can be utilized for the correlation process between transmitter and receiver~\cite{cox1973spatial}. 

This method is used in the establishment of a real-time measurement platform operating at 140~GHz at the New York University~\cite{measureCorCamp1}. This platform is utilized for reflection and scattering propagation characteristics in~\cite{measureCorCamp2}, large-scale fading and multipath characteristics of indoor channels in~\cite{measureCorCamp3}, as well as analyses of receive powers and interference  for  satellite measurements in~\cite{measureCorCampAdd4}. Another correlation based measurement system is developed for 300~GHz at the Technische Universit\"at Braunschweig in~\cite{measureCorCamp4}. They cooperated with Beijing Jiao Tong University (BJTU) and conducted measurements of 60-300GHz channels utilized for China high-speed rail in~\cite{measureCorCamp5, measureCorCampAdd2} with  the train-to-infrastructure inside-station channels~\cite{measureCorCampAdd1} and intra-wagon channels~\cite{measureCorCampAdd3}. The same setup is utilized  for channel measurements of  single-lane and multi-lane vehicle-to-vehicle communications in~\cite{measureCorCamp6} and attenuation by the window and fuselage of a Boeing 737 aircraft~\cite{measureCorCamp7}.

\subsubsection{Direct Pulse Method}

In parallel to the above two approaches, the direct pulse method, also known as THz time-domain spectroscopy (THz-TDS), is the most straightforward, which trains very narrow pulses in the time domain with a period larger than the maximum excess delay (i.e., the relative delay between the first-arriving component and the multipath component whose energy falls below a certain threshold from the strongest one).
Using this measurement method, the channel impulse response can be easily obtained from  the observation of the received signals. This can be realized by considering the amplitude of the channel impulse response as well as the delay between the transmission time and sampling time.
This method is suitable for the THz waves with ultra-broad bandwidth~\cite{measureDP}, while incurring low power, large equipment size, short distance, as well as the range of applications limited to the measurements of reflection, scattering, and diffraction properties induced by the narrow beams.

We have used the THz-TDS method to measure the interference of THz channels at 300 GHz and analyzed stochastically~\cite{measureDPCamp1}.
Measurements have been conducted to study reflection coefficients of building materials at 100--1000~GHz~\cite{measureDPCamp2}  and those of stratified building materials~\cite{measureDPCamp3} by teams from Brown University and the Technische Universit\"at Braunschweig. This method has also been applied to measure diffusion coefficients of rough surfaces at 200--400~GHz~\cite{measureDPCamp4}.


\subsection{Channel Modeling Methodologies}
On the basis of THz channel measurements, channel characterization is needed to support subsequent research on the PHY layer and above, whereas existing channel models for lower frequencies are not directly applicable to the THz band, thus motivating different channel modeling methodologies. Current channel characterization can be performed deterministically, stochastically, or in a hybrid manner combining both synergistically.

\subsubsection{Deterministic Channel Modeling}
\label{sec:SISO-deterministic}
Starting from the EM wave theory~\cite{modelAddEM}, deterministic channel models provide an accurate description of wave propagation.
Numerical results generally conform to the measurements, while the complete geometric information of the wireless environment, attributes of materials, and exact locations of the transceivers are indispensable.
Specifically, we survey two commonly utilized  schemes in the context of THz communications.

\begin{itemize}
    \item \textbf{Ray-tracing (RT)} is a promising choice in modeling site-specific wireless environments and keeps reasonable computational complexity in very large systems. In particular, a fixed transmitter generates signals that encounter different objects with various surface conditions along propagation. Upon reception, the reflected, diffracted, or scattered arrival rays beside the LoS one are captured. By analyzing the wavefronts of each ray as simple particles, the RT explores characteristics of these electromagnetic waves. 
In practice, RT can be realized over several methods. 
A typical method is representing the transmitter as the root node and recursively seeking branches (i.e., objects in the wireless environment that can be connected to this root node in a LoS manner), until the receiver branch is reached. In this way, the RT develops a visibility tree of deducing paths through the backtracking methods.
The other approach is a strategy called ray launching, which radiates rays in a predetermined grid of directions and explores the trajectory~\cite{rayLaughching}.


Since THz signals have stronger corpuscular properties compared to low-frequency bands, geometric optics can be used to accurately simulate the loss in reflection, diffraction, and scattering~\cite{han2018propagation}. 
Recently, in consideration of the suitability of the RT for THz waves, tailored RT modeling for the THz band has been conducted.
For example, the RT method is invoked at 300 GHz in an indoor scenario~\cite{RTcali}, which is calibrated by channel measurements to improve the accuracy.
In addition, we exploit the RT method to generic multipath THz channel models that are applicable to the entire THz spectrum~\cite{Han2015multiray}. 
Besides, by accounting for the elevation plane apart from the conventionally considered azimuth plane, we have successfully applied the RT method to build 3D end-to-end THz channel models~\cite{han2017three,nie2017three}. 
We also discuss Several available THz channel simulation platforms that build on RT theory are avaible as well (discussed in Sec.~\ref{sec_testbeds}). 


\item In parallel to the RT, \textbf{the finite-difference time-domain (FDTD) or Yee's method}~\cite{zhao2007fdtd} numerically derives the channel model from  Maxwell's equations.
Specifically, to accelerate the computation, the whole space is divided into grids that are separately explored. 
Then, both the electric and magnetic fields are analyzed and iteratively updated in the temporal and spatial domains. 
Considering the rigorous theoretical computation, the FDTD can model small and complex scatterers and rough surfaces with high accuracy.
However, the high accuracy is achieved at the expense of a substantial computational burden in  both memory and time in attaining all geometrical features along propagation.

\end{itemize}


\subsubsection{Statistical Channel Modeling}
\label{sec:SISO-statistical}
Based on large datasets obtained from empirical channel measurements, statistical channel modeling (SCM) method has an advantage of low computational complexity compared to the aforementioned deterministic modeling alternative. Since the THz channel is deemed spatially sparse with resolvable multipath components, tailored models are necessary in lieu of narrowband models, in which Rayleigh or Rician fading is commonly assumed.  

Some SCM models characterize the temporal and spatial channel characteristics with fitted distributions, such as directions of departure/arrival (DoD/DoA), time of arrival (ToA), as well as complex amplitudes. The Saleh-Valenzuela (S-V) model with the tapped delay line~\cite{stModelSV} can be used to describe clusters of multipath components. An example of such a model at 300~GHz can be found in~\cite{stModelSVTHz}.
By contrast, some SCM models describe the impulse response of channel and antenna characteristics mathematically without the explicit consideration of propagation. 
On the one hand, the simple assumption of independent and identically distributed (i.i.d.) Rayleigh fading channel model is widely studied.
Nevertheless, the i.i.d. model is not suitable for the THz band due to the high sparsity.
On the other hand, the Kronecker model considering correlation between sub-channels is conventionally utilized as well.
However, in UM-MIMO channels, the Kronecker model can lead to high bias~\cite{han2018propagation}.
To address those challenges, virtual channel representation (VCR) is proposed to explore the beam space when sampling rays for UM-MIMO systems~\cite{synPerBeam}, which however only focuses on uniform linear arrays.
Furthermore, our previous work illustrates that the integration of the Kronecker and VCR models, namely, the Weichselberger model, is valid for various UM-MIMO scenarios~\cite{han2018propagation}.

\subsubsection{Hybrid Channel Modeling}

\label{sec:SISO-hybrid}
As described in Secs.~\ref{sec:SISO-deterministic} and \ref{sec:SISO-statistical}, deterministic channel modeling achieves high accuracy at the expense of low time and resource efficiency.
By contrast, statistical channel modeling works with a low computational burden while sacrificing accuracy. 
Thus, by strategically fusing two or more approaches, the development of \textit{hybrid methods} is attractive and promising to resolve complex THz channel models under low latency.

For instance, a combination of RT and FDTD methods can realize accurate modeling in a time-efficient manner~\cite{RT_FDTD}. 
Specifically, the FDTD with high accuracy assists the modeling in the region close to scatterers, while the RT is in charge of modeling the rest of the wireless channel.
This hybrid method has been utilized in~\cite{RT_FDTD_More} for indoor channel modeling.
Compared to the individual methods, this hybrid scheme is preferred in THz communications due to its high time efficiency and accuracy.

Furthermore, the hybrid approach can provide even higher computation efficiency owing to the stochastic component.
Sufficient channel information can  also be captured for ease of higher accuracy thanks to the deterministic part.
With the DoA following a Gaussian mixture model (GMM), this hybrid modeling has been implemented at 0.3~THz for an indoor scenario~\cite{choi2013geometric}.
By generating angles of arrival (AoAs) with Von Mises distributions, we have established a semi-deterministic THz channel model in~\cite{chen2021channel}.
Moreover, a generic 3D THz channel model under non-stationarity is proposed, which jointly considers space, time, and frequency domains~\cite{hyModel3D}.

\subsection{Channel and Propagation Characteristics}
\label{sec:channel char}
In addition to the aforementioned modeling techniques, a comprehensive understanding of fundamental spectral and spatial characteristics is critical. We hereby summarize them in Table~\ref{tab:channel_summary} and elaborate as follows.
\begin{enumerate}
	
\item\textit{Spreading Loss:} The Friis' law with frequency-independent antenna gains suggests that when the antenna apertures at both  link ends decrease, the spreading loss increases quadratically.
However, fixed-aperture antennas with frequency-dependent gains will leads to a quadratic  drop of the  loss.

\item\textit{Atmospheric Loss:}
The atmospheric loss or molecular absorption loss is caused by the partial conversion of THz wave energy into a rotational transition form for polar gas molecules in the propagation medium. Such molecules are usually composed of water vapor and oxygen molecules.
As a result, the power of the THz signal is attenuated, while the noise is amplified~\cite{chaccour2021seven}.
Moreover, the absorption peaks create spectral windows in the THz band, which are  distance-varying~\cite{han2018propagation}.
For example, as shown in Fig.~\ref{fig:sw_los}, two windows at 0.38--0.4 and 0.62--0.72 THz with a 30~m transmission distance merge into one at a distance of 1 m.
Compared with other spectrum bands used for wireless communications, the atmospheric loss is relatively small for both mm-wave (caused by oxygen molecules) and OWC communications (owing to water vapor and carbon dioxide molecules), while negligible in microwave communications.
Such a noticeable atmospheric loss in THz band requires physical layer designs with distance adaptivity and multi-wideband capability~\cite{han2016distance}.

\begin{figure*}[t]
\centering
\includegraphics[width=0.8\textwidth]{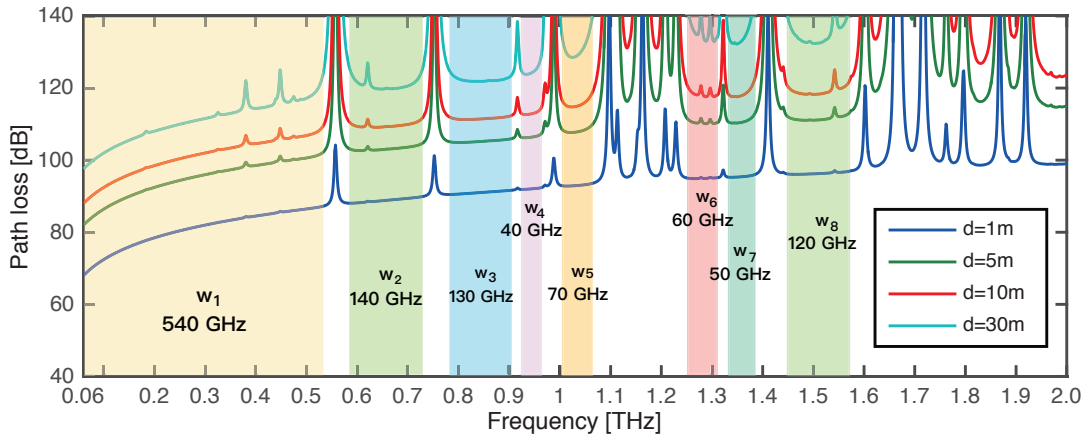} 
        \caption{Path loss in the THz band. Eight spectral windows are identified and their bandwidths range from 40 GHz up to 540 GHz.}
        \label{fig:sw_los}
\end{figure*}

\item \textit{Temporal Broadening:} 
Besides spectral windows, the frequency-selectivity of THz communications incurs broadening in the temporal domain. 
By defining the \textit{broadening factor} as the quotient of the durations for the received and transmitted pulses, the minimal interval between consecutive transmissions without inter-symbol interference can be quantified.
The value of this factor increases with growing signal bandwidth, operating frequency, and transmission distance~\cite{Han2015multiray}, hence limiting the transmission rate.

\item\textit{Diffuse Scattering and Specular Reflection:}
The comparable scales of wavelength in the THz band and the surface roughness of common objects incur severe diffuse scattering on rough surfaces, where scattered rays are radiated to various directions. 

The loss stemmed from specular reflection is a function of the electrical thickness of the surface, which is a frequency-dependent parameter.  
With respect to all multi-paths, the received power from specular reflection is generally dominant as compared to that of diffuse scattering.
Moreover, as the frequency increases, the surface roughness parameter that determines the scattered power declines, while the surface correlation parameter controlling the scattered beam width increases.
In the sequel, the rays at very high frequencies travel in all directions with uniform power after encountering severely rough surfaces.

\item\textit{Diffraction, Shadowing, and LoS Probability:}
Due to the sharp shadows of various objects, such as walls and people, the efficiency of diffraction decreases with increased frequency. Hence, in practice, diffraction can be neglected  so long as the receiver is in a very closed region approaching the incident shadow boundary.
As a consequence, as the frequency increases, shadow fading variance becomes more significant.
Moreover, the area within the first Fresnel zone that is especially vulnerable to shadowing objects reduces with the square root of the wavelength.
In addition, generally, the frequency does not impact the LoS probability.
However, owing to the skin effect in lossy media, transmission power decreases near uniformly with increasing frequency.

\item \textit{Coherence Bandwidth:} 
Within the range of coherence bandwidth that is inversely proportional to the delay spread, the signal can be transmitted without undergoing the frequency-selective fading. 
In the THz band, the larger propagation distance induces the narrower coherence bandwidth~\cite{Han2015multiray}. {Between 0.1 and 1~THz, the average coherence bandwidths are 3.42 GHz, to 2.56 GHz, 2.17 GHz, 1.75 GHz, and 1.47 GHz, by altering the distances as 2 m, to 3 m, 4 m, 5 m, and 6 m, respectively. In general, an increasing frequency would lead to multipath effect dwindling, due to the very high loss for the NLoS paths. This results in a smaller delay spread and thus, a larger coherence bandwidth.}
In addition, our beamforming design elongates the THz signals for farther while slenderer coverage, resulting in a reduction for {multi-path component (MPC)} and thereby a wider coherence bandwidth~\cite{han2017three}.

\item \textit{Stationarity Time and Stationarity Bandwidth:} 
Stationarity  time and bandwidth are defined as the period and the spectrum range that emerges  wide-sense stationary channels in time-varying THz communications.
Compared with the coherence region~\cite{chen2019time}, the stationary region in THz vehicle-to-everything (V2X) communications is much larger, since the stationarity region depends on the statistical channel parameters, while the coherence region is determined by the exact values for these parameters. 
Furthermore, the shorter stationarity distances suggest the heavier non-stationarity in THz V2X channels.
Experiments demonstrate the approximate 0.37~m stationarity distance when the receiver moves with a velocity of 15~m/s, revealing the small variation of the approximate wide-sense stationary environment within 0.37~m.

\item \textit{3D Angular Spread:} 
As a key to spatially characterizing the 3D channels, especially those equipped with UM-MIMO, the angular spreads need to be studied in the elevation plane as well as the azimuth plane.
According to the trials in our previous work, the beamforming and the relative locations between the transmitter and receiver remarkably affect the angular spreads~\cite{han2017three}.

\item \textit{Spatial Degree of Freedom:}
The spatial degree of freedom depends on the array area, array geometry, and angular spread. This parameter measures the largest spatial multiplexing gain that can be supported by a UM-MIMO system~\cite{han2018ultra}. 
Therefore, the large angular spread in a rich scattering environment leads to a sizable spatial degree of freedom. 
\item\textit{Weather Influences:}
The aerial particulates, including rain, fog, haze, smog, etc., further increase the THz  propagation loss.
This property can be characterized via the Mie scattering theory when airborne particles are in the comparable size of the THz wavelength.
Otherwise, with smaller airborne particle sizes, the Rayleigh model is used to approximate the Mie scattering model.
In addition, when the sky is not clear, considerable attenuation occurs on the infra-red or OWC rays whose  wavelengths are smaller. 
However, better robustness against weather influences emerges at the THz band, since the THz signals are less susceptible when traveling through small airborne particulates.

\item\textit{Scintillation Effects:}
The presence of thermal turbulence near the ground incurs the spatial and temporal inhomogeneities of temperature and pressure across time and space in the air.
As a result, scintillation can arise.
In particular, the various refractive indices for the wavefront of a beam and unevenly distributed  deflection and interference lead to the distortion of the flat phase front of the beam.
Owing to this effect, a twinkling speckle pattern, which contains severe temporal variations on intensity, occurs in the beam cross-section at the receiver.
In comparison, the THz communications are immune to the effects of scintillation, while these effects confine the largest distance of OWC communications.
\end{enumerate}

\begin{table*}
\centering
\caption{THz channel features and impact on 6G wireless communication and networking.}
\label{tab:channel_summary}
\resizebox{\textwidth}{!}{
\begin{tabular}{|m{3cm}<{\centering}|m{4.2cm}<{\centering}|m{4.2cm}<{\centering}|m{4.2cm}<{\centering}|}
\hline
\textbf{Parameter}                                   & \textbf{Dependence on frequency}                                                                                                & \textbf{Impact on 6G THz systems}                                              & \textbf{THz vs. Microwave and OWC}                                                                                                                          \\ \hline
Spreading Loss                                       & Quadratic increase with decreasing area and constant gains; Quadratic decrease with constant area and frequency-dependent gains & Distance limitation                                                            & Higher than microwave, lower than OWC                                                                                                                       \\ \hline
Atmospheric Loss                                     & Frequency-dependent path loss peaks appear                                                                                      & Frequency-dependent spectral windows with varying bandwidth                    & No clear effect at microwave frequencies, oxygen molecules at millimeter wave, water and oxygen molecules at THz, water and carbon dioxide molecules at OWC \\ \hline
Temporal Broadening                                  & Increase with bandwidth, frequency, and distance                                                                                & Limited minimal spacing between consecutive pulses and data rate               & Stronger than microwave, weaker than OWC                                                                                                                    \\ \hline
Diffuse Scattering and Specular Reflection           & Diffuse scattering increases with frequency. Specular reflection loss is frequency-dependent                                    & Limited multi-path and high sparsity                                           & Stronger than microwave, weaker than OWC                                                                                                                    \\ \hline
Diffraction, Shadowing and Line-of-Sight Probability & Negligible diffraction. Shadowing and penetration losses increase with frequency. Frequency-independent LoS probability         & Limited multi-path, high sparsity and dense spatial reuse                      & Stronger than microwave, weaker than OWC                                                                                                                    \\ \hline
Coherence Bandwidth                                  & Decrease with longer distances and constant numbers of MPCs; increase with constant distances and fewer numbers of MPCs         & Limited data rate without the frequency-selective fading                       & Broader than microwave, narrower than OWC                                                                                                                   \\ \hline
Stationarity Time and Stationarity Bandwidth         & Decrease with frequency                                                                                                         & Constraint in mobile system design                                             & Smaller than microwave, larger than OWC                                                                                                                     \\ \hline
3D Angular Spread                                    & Decrease with frequency                                                                                                         & Constaint in beamforming and beam tracking design                              & Smaller than microwave, larger than OWC                                                                                                                     \\ \hline
Spatial Degree of Freedom                            & Decrease with frequency                                                                                                         & Constaint in beamforming and beam tracking design                              & Smaller than microwave, larger than OWC                                                                                                                     \\ \hline
Weather Influences                                   & Frequency-dependent airborne particulates scattering                                                                            & Potential constraint in THz outdoor communications with heavy rain attenuation & Smaller than microwave, larger than OWC                                                                                                                     \\ \hline
Scintillation Effects                                & Increase with frequency                                                                                                         & Constraint in THz space communications                                         & No clear effects at microwave, THz is less susceptible than OWC                                                                                             \\ \hline
\end{tabular}
}
\end{table*}

\subsection{Link Budget Analysis}
\label{sec:link_budget}

To support the realistic commercialization applications, the effectiveness and reliability of THz communication systems should be verified. 
To this end,  practical link budget indicators need to be thoroughly evaluated in the THz band, including frequency, desired data rate noise figure, etc. 
Two fundamental knowledge are required to calculate the THz link budget.
In particular, on one hand, theoretical THz propagation and channel characteristics provide prerequisite calculus terms for link budget, which are discussed in Sec.~\ref{sec:channel char}.
On the other hand, numerous practical measurement campaigns elaborated in Sec.~\ref{channel measurement} can be a good guideline for the link budget analysis.
For example, we list the THz link budget for an indoor scenario in Table \ref{tab:link budget}.
More specifically, to achieve 1~Tbps downlink, 1,024 Tx antennas, 64 Rx antennas, and 4 data streams are deployed, and the link distance is 1 meter.
Note that these parameters are adjusted according to the simulation with the noise figure computed by following the scheme in~\cite{rikkinen2020thz}. 
If the desired data rates are 100~Gbps at the uplink and 200~Gbps at the downlink, a transmission distance longer than 10~m can be achieved.

We conduct another analysis investigates the link budget parameters for AR/VR-fronthauling as well as backhauling in~\cite{linkBudge1}, to provide a quantitatively perspective on such systems at 287.28~GHz. 
Specifically, for AR/VR over 10~m, the achievable rate reaches 20~Gbps with 10~GHz bandwidth and 10~dBm transmit power. By contrast, the rate for backhauling achieves 200~Gbps, when the bandwidth and transmit power increase to 100~GHz and 20~dBm, respectively.
The distance in this scenario can extend up to 2~km if 55~dBi antenna gain is available.


\begin{table}
	\centering
	\caption{THz link budget for different scenarios.}
	\label{tab:link budget}
	\begin{tabular}{|c|m{2cm}<{\centering}|m{2cm}<{\centering}|}
\hline
\textbf{Parameter}     & \textbf{Uplink}   & \textbf{Downlink} \\ \hline
Frequency              & \multicolumn{2}{c|}{300 GHz}          \\ \hline
Bandwidth              & \multicolumn{2}{c|}{30 GHz}           \\ \hline
Target achievable rate & \multicolumn{2}{c|}{1 Tbps}           \\ \hline
Tx beamforming gain    & 18.1 dBi          & 30.1 dBi          \\ \hline
Rx beamforming gain    & 30.1 dBi          & 18.1 dBi          \\ \hline
Transmit power         & 5 dBm             & 13 dBm            \\ \hline
Link distance          & \multicolumn{2}{c|}{1 to 100 m}       \\ \hline
Rx noise figure        & \multicolumn{2}{c|}{10 dB}            \\ \hline
Number of Tx antennas  & 64                & 1024              \\ \hline
Number of Rx antennas  & 1024              & 64                \\ \hline
Antenna gain           & \multicolumn{2}{c|}{5 dBi}            \\ \hline
Achievable Rate        & 1171 down to 12.8 Gbps & 1490 down to 50.1 Gbps \\ \hline
\end{tabular}
\end{table}

\subsection{Open Challenges}

\begin{itemize}
	\item \textbf{Efficient THz Channel Measurement Systems}.
	Most of the existing measurement methods are adapted from systems designed for low-frequency bands, i.e., mm-wave.
	However, as we mentioned in Sec.~\ref{sec:channel char}, the special properties of THz waves motivate more advanced and flexible systems to handle significant path loss, ultra-large bandwidth with high frequency, as well as short coherence time.
	Based on such systems, extensive channel measurement campaigns should be conducted for various scenarios, such as unmanned aerial vehicles assisted wireless communications, virtual/augmented/mixed/extended reality (VR/AR/MR/XR), nano-communications, etc., for ease of the practical implementations of these THz applications.
	

	\item \textbf{Accurate and Flexible THz Channel Models}.
	The discussions in   Sec.~\ref{sec:SISO-deterministic}  reveal the shortage of the commonly used deterministic and stochastic channel models for THz communications. These issues motivate the integration of deterministic and statistical modeling approaches to reach both high efficiency and accuracy under multi-paths.
	However, the smooth transition between different modeling schemes, the requirement of substantial measurement campaigns, and the complex parameter extraction is still a challenge. The design for accurate and flexible hybrid models is thereby limited.
	Moreover, considering the promising prospect of THz UM-MIMO systems with high spectrum and power efficiency, the high-performance THz UM-MIMO channel modeling is significant for next-generation wireless communications. 
	However, the large-scale antenna arrays exploited by UM-MIMO systems lead to multiple issues that need to be innovatively addressed for modeling, including the complex analysis and mutual coupling effect.  
	
	\item \textbf{Outdoor and Mobile Channel Measurements and Models}. 
	Recently, some THz channel measurements and modeling solutions have been for several outdoor and mobile scenarios, including vehicular networks~\cite{measureCorCamp6}, terrestrial and satellite spectrum sharing systems~\cite{measureCorCampAdd4}, airplane-satellite~\cite{kokkoniemi2021channel}, as well as inter-satellite links~\cite{nie2021channelsatellite}. 
	This suggests that the outdoor and mobile THz channels have drawn remarkable research attention, while further investigation of non-stationary and outdoor scenarios still remain open issues.
	
	\item \textbf{AI-enabled Channel Modeling}.
	Thanks to the sustaining advancement of AI techniques, AI-based wireless communication has immense potential to address complicated problems.
	{Attempts have been conducted to utilize AI in channel modeling, in aspects of channel parameters analysis~\cite{he2018clustering}, as well as path loss prediction~\cite{ostlin2010macrocell}.}
	Hence, taking into account the special THz properties together with the requirements of accurate multi-path channel measurement analysis and modeling, AI techniques are preferred by the complex THz systems.

\end{itemize}

\section{PHY Layer Functionalities}
\label{sec_PHY}

In light of the ongoing THz hardware and channel developments, efficient PHY layer technologies are needed to meet the constantly increasing system and network requirements.
Till date, the hardware implementations are still rare, and only a few of the works discussed in this section have been in fact experimentally validated. Promisingly, diverse theoretical designs have been proposed, which are potential to realize efficient communication in the THz band.
In this section, we focus on innovative technologies at the physical layer.

\subsection{Modulation and Coding Schemes}
\label{sec_modulation}
Modulation and coding are the primary elements for reliable wireless communication designs.
Considering unique device and channel properties in the THz band, classic modulation and coding schemes need to be revisited while new ideas to be developed.
When we started exploring the physical layer of THz communications in 2010~\cite{Akyildiz,teranets}, we focused on \textbf{short-range communications}. 
In such cases, ultra-low transmission power and low-complexity implementation are required. 
As discussed in Sec.~\ref{sec_channel}, for very short distances, molecular absorption loss is almost negligible. 
Therefore, the THz band exhibits itself as almost a 10-THz-wide window. 
To make the most out of this bandwidth with minimal complexity, we proposed the utilization of one-hundred-femtosecond-long pulses.
Such pulses are transmitted by following an on-off keying modulation and spreading the symbols in time (TS-OOK)~\cite{jornet2011information,TSOOK}. 
In TS-OOK, 1s and 0s are transmitted as pulses and silence, respectively, which drastically simplifies the transmitter and the receiver architecture. 
Moreover, TS-OOK aims to further relax the hardware requirements and enable uncoordinated multiplexing of parallel streams between different users. 
To this end, instead of being sent back to back, symbols are spaced out in time to avoid temporal broadening effects (as elaborated in Sec. IV-C) and allow for multiple access. 
{Indeed, TS-OOK is listed in the IEEE 802.15.3d standard, which just detect binary information~\cite{IEEE802_15_3d}.}
{
{In addition, benefiting from the single option of OOK modulation scheme defined in the IEEE 802.15.3d standard, fewer bits are required in the PHY overhead, compared to the signal carrier mode that supports multiple options of modulation (e.g., phase-shift keying (PSK) and quadrature amplitude modulation (QAM)).}}
Besides the TS-OOK reducing the PHY overhead,  an index modulation has been applied to assist the THz pilot design with shorter codewords~\cite{mao2021terahertz}. 
Specifically, this method maps the indices of data and pilot subblocks onto short index bits with the logarithmic length.

When increasing the communication distance to \textbf{macro-/micro-scale}, the distance-dependent bandwidth of the THz band created by molecular absorption needs to be accounted for. 
In this direction, in~\cite{Han2016wideband}, we proposed the utilization of multiple transmission windows, each on its turn with a variable number of narrower-band pulses. 
More specifically, by ensuring the data rate demands, the optimal waveform design is proposed to determine the transmit power and sub-window rate. 
THz communication distances can thereby be maximized with thorough consideration of multiple peculiarities of THz waves, including the distance-varying spectral windows, the delay spread, as well as the temporal broadening effects.
More recently, in~\cite{hossain2019hierarchical}, we have introduced the concept of hierarchical bandwidth modulations (HBMs). The goal of HBM is to simultaneously transmit independent data streams to multiple users at different distances within the same transmission beam.
To this end, HBMs dynamically exploit the distance-dependent bandwidth of the THz band and multiple modulation constellations of all the users.
More specifically, HBMs build on top of traditional hierarchical or concatenated modulations~\cite{jiang2005hierarchical}.
Users are able to recover different information based on their perceived signal-to-noise ratio (SNR).
As a result, users at different distances can utilize a different bandwidth.
The atmospheric loss issue is hence turned into an opportunity of the utilization maximization for distance-dependent THz bandwidth. 




Innovative waveforms are needed to support \textbf{ISAC}~\cite{wu2021ISCI}. Several attempts have been recently reported.
First, we have applied non-uniform subcarrier spacing and superposition to multiple orthogonal frequency division multiplexing (OFDM) waveforms, and the sensing accuracy is dramatically improved, and the throughput is enhanced~\cite{sensingNonUniform}.
Besides multi-carrier design, we have presented a single-carrier study in~\cite{sensingDftOFDM}, in which we proposed a sensing integrated discrete Fourier transform spread orthogonal frequency division multiplexing (SI-DFT-s-OFDM) to explore the distinct peculiarities of sensing and communication channels.
Simulation results illustrate that SI-DFT-s-OFDM realizes ten times higher estimation accuracy, compared to the conventional OFDM. 
Third, a new modulation technique named orthogonal time frequency space (OTFS) was recently proposed to address high Doppler spread in doubly selective channels~\cite{wei2021orthogonal}. OTFS maps the information symbols on the delay-Doppler domain, which provides a sparse representation of a time-varying multi-path channel. To unleash the potential in the THz band, we have developed a discrete Fourier transform spread OTFS (DFT-s-OTFS) modulation scheme for THz communications in~\cite{wu2021dftsotfs}, which can achieve strong robustness to Doppler spread compared to OFDM/DFT-s-OFDM and provide lower peak-to-average power ratio (PAPR) in contrast with OTFS. This design could be further explored to include the sensing function.


In addition to novel modulation schemes, customized coding designs for THz communications are proposed.
    {For instance, as discussed in the IEEE 802.15.3d standard~\cite{IEEE802_15_3d}}, the header check sequence (HCS) is exploited to ensure the correctness of the PHY and MAC headers for THz signals. The robustness is enhanced by encoding HCS together with these headers through the extended Hamming code.
    Pertaining to the OOK modulation, IEEE 802.15.3d suggests the Reed Solomon code to realize hard-decision yet simple binary decoding in addition to the low-density parity-check (LDPC)-based methods.

Moreover, effective next-generation forward error correction (FEC) schemes are required to resolve the channel errors in the THz band. 
Motivated by this, it is essential to know the nature of such errors by developing stochastic models of noise, multi-path fading, and interference first.
Then, with the knowledge of errors, advanced state-of-the-art error control policies such as turbo, LDPC, and polar codes can be the potential candidates to be implemented in the THz band.
Rather than adopting existing coding schemes, ultra-low-complexity channel coding schemes are also favored as excellent solutions for THz communications. 
More specifically, such schemes can preclude channel errors instead of correcting them afterward.
For example, for short-range communications, with femtosecond-long pulse-based modulations, we have successfully applied low-weight coding schemes~\cite{jornet2014low,yao2019ecp}.
In the future, these FEC technologies need further validation and demonstration on a practical THz chipset. 

\subsection{Dynamic Hybrid Beamforming}
\label{DAoSA}
As discussed in Sec.~\ref{sec_device}, owing to the sub-millimeter wavelength of THz signals, very-large-scale antenna arrays  can be applied.
For instance, 1024, 4096, and even more antenna elements can be integrated on the antenna arrays for a transceiver.
As a result, the promising UM-MIMO systems can be achieved.
On one hand, such systems can achieve high beamforming gains, which combat the short transmission distance problem incurred by the high path loss.
On the other hand, multiplexing gains can be realized to further improve spectral efficiency by transmitting multiple data streams simultaneously.

\begin{figure}
	\centering
	\includegraphics[width=\linewidth]{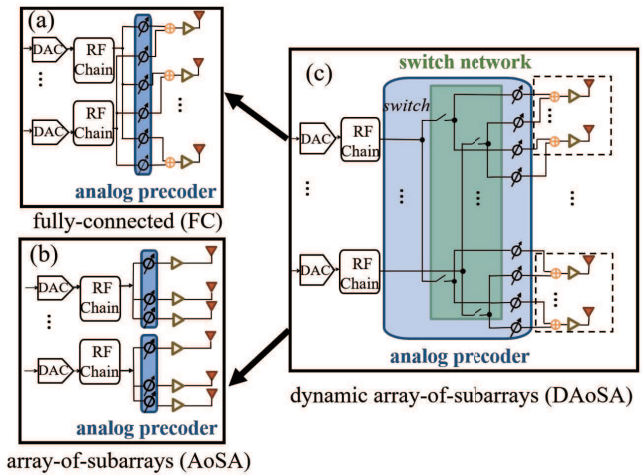} 
	\caption{Three hybrid beamforming architectures for THz communications, namely, (a) fully-connected (FC), (b) array-of-subarray (AoSA), and (c) dynamic AoSA}
	\label{fig:beamforming}
\end{figure}

Instead of conventional analog and digital beamforming architectures that are limited by performance and hardware constraints, hybrid beamforming is appealing in the THz band~\cite{Peng2019precoding, yuan2020hybrid,gao2021wideband,wan2021hybrid,hao2021robust,ning2021terahertz, han2021hybrid}.
Specifically, in the hybrid structure, signal processing is divided into a digital baseband and {an RF} part through a phase shifter network.
As illustrated in Fig.~\ref{fig:beamforming}, hybrid beamforming can be further classified into three
categories, namely, fully-connected, array-of-subarray (AoSA), and dynamic AoSA architectures.
As dynamically structured hybrid beamforming, the third architecture proposed in our previous works can reach a balance among good spectrum efficiency, low hardware complexity, and energy efficiency~\cite{yan2020hybrid,yuan2022cluster}.
Moreover, this dynamic structure intelligently customizes the connections between RF chains and subarrays via a network of switches, to evolve as the fully-connected and AoSA counterparts. 
To realize this, on the one hand, each of subarrays connects to a unique RF chain through a set of switches. 
On the other hand, the authors in~\cite{HBF_combiner} propose to connect each subarray with multiple RF chains via a combiner. 

\begin{figure}
	\centering
	\includegraphics[width=\linewidth]{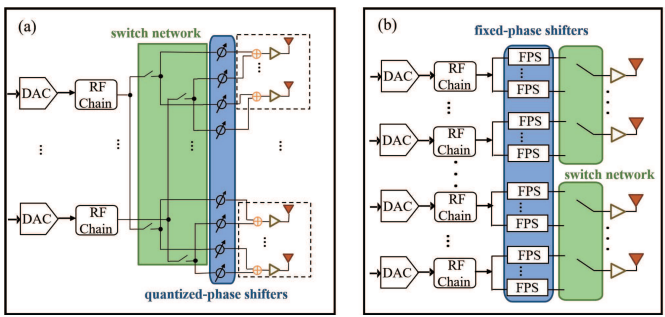} 
	\caption{(a) Dynamic hybrid beamforming architecture with quantized-phase shifters, (b) Dynamic hybrid beamforming architecture with fixed-phase shifters.} 
	\label{fig:hbf_q_f}
\end{figure}

Practical design has been recently proposed to adopt quantized phase shifters and even fixed phase shifters to replace the infinite-resolution phase shifters that are costly with high hardware complexity.
As shown in Fig.~\ref{fig:hbf_q_f}, we have devised the dynamic hybrid beamforming with quantized phase shifters that operate discrete phases~\cite{DAOSA_q_f}.
This system realizes 36\% energy efficiency enhancement at the expense of only 2\% spectral efficiency, compared to the scheme with infinite-resolution phase shifters.
Moreover, by equipping with more economical fixed-phase shifters, the dynamic hybrid beamforming can further achieve 30\% more energy efficiency than that with quantized-phase shifters~\cite{DAOSA_q_f}. 
However, 21\% reduction in terms of the spectral efficiency comes at the price.

In the THz UM-MIMO systems, with broad bandwidth and very large antenna array, a beam squint problem arises since the phase shifter cannot generate frequency-proportional weights. Fortunately, the beamforming weights generated by true-time-delay (TTD) is frequency-proportional and can be used to solve beam squint. Hence, hybrid beamforming using TTD for phase adjustment is a promising alternative to address the wideband beam squint problem in the THz band, while offering high array gains~\cite{HBF_delay}. As one step further, we propose a novel dynamic-subarray with fixed true-time-delay (DS-FTTD) architecture in~\cite{yan2022energyefficient}, which owns lower power consumption and hardware complexity, thanks to the low-cost FTTDs.

\subsection{Beam Estimation and Tracking}
\label{sec_beamalignment}

As a key of beam management, beam alignment that involves angle estimation and tracking is critical. Mainly,  \textit{on-grid} and \textit{off-grid} methods are categorized in the literature for beam estimation and tracking. 
On the one hand, on-grid methods utilize predetermined spatial grids to estimate the DoA, for which the accuracy is thereby dependent on the grid resolution~\cite{alkhateeb2014channel,lin2017subarray}.
As downside, to reach the desired resolution, the on-grid schemes suffer from high complexity and large beamforming training (BFT) overhead.
On the other hand, off-grid methods perform subspace-based super-resolution DoA estimation or discard the fixed angular grids.
To support the super-resolution DoA estimation for the aforementioned subarray-based hybrid beamforming structure, we have proposed a THz-specific AoSA-multiple signal classification~(AoSA-MUSIC) in~\cite{beamTrackingMUSIC}.
Simulation results demonstrate its ability to realize millidegree-level angle estimation and millisecond level tracking.
In addition, a variant of subspace pursuit (SP) algorithms in compressed sensing (CS) is also utilized to estimate the angle of beams with low computational complexity for THz MIMO systems~\cite{ma2020joint}.
Moreover, a major challenge of beam estimation relates to the beam squint problem, or the so-called triple delay-beam-Doppler squint effects. 
To tackle this, a variant of the traditional orthogonal matching pursuit (OMP) is exploited to solve the CS problem~\cite{dovelos2021channel}.
Moreover, the authors in~\cite{liao2021terahertz} study the estimation and tracking of fine beam angles in space-air-ground networks.
To further ensure the beam alignment in 3D THz scenarios, we propose a low-complexity beam training and tracking approach in~\cite{ning2021unified}.
Benefiting from a hierarchical codebook including narrow beams and wide beams, this method can enhance the worst-case performance as well as reduce the dead zone, respectively.


Besides the above methods, an innovative beamforming architecture that utilizes the beam squint effect can, in turn, inspire the design of beam estimation and tracking.
When transmitting THz wideband signals, the time delay elements can also make phase shift frequency-dependently, which is called THzPrism~\cite{THzPrism}.
This property makes the phased array radiate different frequency carriers to diverse directions and remarkably broaden the angular coverage.
For example, THzPrism-based hybrid beamforming achieves 124\% more data rate, serves 147\% more users, while consuming 71\% less power than the existing approaches~\cite{zhai2021ss}.
With such a structure, multiple beams can be generated and utilized simultaneously for tracking, based on the pairing between frequency and beam angle~\cite{THzPrismTracking}.
Specifically, accurate tracking results are obtained with 90\% less overhead, compared to the counterpart hybrid beamforming architectures.

Thanks to the recent rapid development of DL for wireless communications, inherent attributes of channels such as angle information can be excavated via DL-based methods.
For example, we have deployed a deep convolutional neural network (DCNN) in THz networks, and the beam estimation is enhanced in terms of both accuracy and complexity~\cite{beamTrackingDL1}.
Besides, we have further developed DCNN as well as the convolutional long short-term memory (ConvLSTM) for beam estimation and tracking, respectively~\cite{beamtrackingMusic2}.
DCNN outperforms non-learning methods under high SNR, while ConvLSTM can realize the accuracy on the order of $0.1^{\circ}$ at a millisecond (ms) level with 50\% reduced overhead.
Thus, DL-based methods are promising to reach higher efficiency and accuracy in complex THz wireless environments, which still needs further research efforts.

\subsection{Synchronization}

According to the Nyquist sampling theorem, sampling the received signal and performing sophisticated digital signal processing are quite challenging to meet Tbps transmission rates.
Novel designs for synchronization is thereby motivated. 
Aiming at realizing the efficient synchronization in THz systems,
our solution determines the symbol start time and shortens the observation window iteratively~\cite{synPulseBase}.
This solution achieves the tight synchronization with a low symbol error rate and a few preamble bits under the pulse-based modulation.
Besides, by jointly exploiting one receiver-initiated handshake mechanism and the aggregated packet or sliding window flow control, we have proposed a synchronization method in~\cite{synPulseAndCarrier}. 
This method is applicable to the carrier-based scenarios apart from the pulse-based modulation.
Furthermore, to realize the sub-Nyquist sampling, we extended the sampling theory with finite rate from compressive sampling to the communication context~\cite{han2017timing}.

In addition to timing acquisition, robust and comprehensive synchronization mechanisms for carrier frequency and phase are needed as well. 
Since the correlation between beam domain channel elements and time/frequency vanishes gradually with more antennas at transceivers in mm-wave/THz MIMO systems, separate synchronization for each beam is developed in~\cite{synPerBeam} to exploit this phenomenon.

\subsection{Localization}
Accurate localization is required for 6G and Beyond wireless systems. With the ultra-broad bandwidth of the THz band, the localization performance is expected to reach 50 centimeter outdoor, 1 centimeter indoor, as well as 1--3 mm imaging resolution~\cite{chen2021cutting,sarieddeen2020next}.
Recent attempts commence on the low-cost THz backscattered systems.
Specifically, in such systems, the reader devices radiate signals, and the passive tag devices modulate and reflect part of the signal back to the reader.
Following this direction, systems based on chipless radio frequency identification (RFID) have been developed to solve the indoor localization with millimeter-level accuracy~\cite{el2018distance,el2018high}.
In particular, the tags with known geographic positions are served as the reference, and the reader predicts its own position on the basis of the backscattered signals.
Besides the indoor localization, we have implemented the THz backscattered communications to localize the in-body bionanosensors~\cite{simonjan2021body}.
In this system, in-body bionanosensors deploy inertial measurement units (IMUs) with nanoscale accelerometers and gyroscopes.
Such units enable the sensors to leverage the inertial positioning, and then the operator collects the position information through the resource-efficient THz backscattered links.

Apart from the backscattered systems, an innovative cooperative localization scheme has been proposed in~\cite{stratidakis2019cooperative} for complex three-dimensional indoor environments, e.g., office and laboratory. 
Particularly, multiple base stations, each of which performs an AoA tracking, cooperate with each other and increase the localization accuracy with low-estimation overhead.
In addition, we have specifically designed recurrent neural networks exploring multiple features, including power, 3D AoA, and ToA, of all the MPCs from multiple transmitters~\cite{fan2020structured,loc}.
This  neural network enhances the accuracy of the indoor localization to centimeter-level in complex indoor environment, which needs further improvement to reach the goal of millimeter-level accuracy. 

\subsection{Physical Layer Security}
Besides the enticing data rate owing to the abundant bandwidth resource, the potential of physical layer security of THz communications needs in-depth analysis.
Indeed, the potential of THz communications for enhancing information security has attracted great attention due to its high directivity and high path loss~\cite{secureSurvey,gao2020receiverGC,gao2019distanceGC}.
Although owing to the high directivity in THz systems, the probability that eavesdroppers locate in the beam sector of the legitimate transmission is low, the security of THz communications is still an essential issue for the following three reasons. 
First, as long as the eavesdropper is inside the beam sector for the legitimate communication, the directivity of THz beams fails to bring security.
Second, signals can be intercepted by an object laying on the LoS paths and scattered towards the eavesdropper~\cite{secureSurvey}.
Third, the multipath-based method or circumventing eavesdropping~\cite{secureMPC1,secureMPC2} can reduce the eavesdropping probability and increase the secrecy rate. 
These reasons motivate the exploration of novel countermeasures of physical layer security.

To start with, security schemes that consider the special attributes of the ultra-spread THz spectrum can be investigated.
To this end, we have devised a distance-adaptive absorption peak modulation (DA-APM) scheme in~\cite{secureDAAPM}.
The main principle is to utilize the the frequency-dependent molecular absorption, and the solution is based on spectrum management, by selecting frequency bands with proper molecular absorption coefficients.
As a result, the eavesdroppable area is largely shrunk although both eavesdropper and receiver reside in the beam sector of the transmitter.
Nevertheless, this scheme is valid only when the eavesdropper is farther than the legitimate receiver.

On the contrary, in the case of eavesdroppers in close proximity where
eavesdroppers are nearer from the transmitter than the legitimate user, a widely-used idea is to simultaneously transmit artificial noise (AN) signals. The AN signal does not contain information, which is sent together with information signals to confuse the eavesdroppers.
Most existing studies have suggested the transmitter to generate AN signals, which nevertheless are not applicable for THz LoS transmission, since the LoS path is shared between the legitimate user and the eavesdroppers. Instead, we have proposed a molecular absorption aided receiver AN scheme in the THz band, where the expensive self-interference cancellation (SIC) is not needed~\cite{secureClose}.
The key idea is to design the information and AN waveforms based on the temporal broadening effect caused by molecular absorption (as discussed in Sec.~\ref{sec:channel char}). 
Specifically, the higher molecular absorption of THz waves at a longer distance can lead to more severe pulse broadening in the time domain.
Based on this, the received signal detection period and constituent ratios of transmission periods (for sending both information signals and ANs) are cooperatively devised.
Consequently, eavesdropping is prevented by the receiver AN scheme, while legitimate communication is guaranteed.

\subsection{Open Challenges}

\begin{itemize}
	\item \textbf{PHY Solutions for Terahertz Integrated Sensing, Communication, and Intelligence (ISCI)}.
	On the one hand, an integration of THz sensing and communication enables simultaneous message transmission and environment perception with shared spectrum, hardware, signal processing modules. On the other hand, driven by ubiquitous signal streams and sensing data, AI can mine the information and further empower the wireless system. 
	PHY solutions for the THz ISCI need to be developed from the following three aspects. 
	First, related to the discussion in Sec.~\ref{sec_modulation}, sensing function needs to be incorporated in the modulation and waveform design, with multiple objectives including sensing targets and information transmission.
	Second, beamforming technologies that balance the tradeoff between communication and sensing need to be developed, due to the distinction of their objectives. That is, sensing prefers scanning beams to search targets, while communication requires stable beams toward receivers.
	Third, AI methods can address the imperfections and non-linear distortion of THz transceivers. As a result, signal detection, sensing parameter estimation, and even an integrated detector based on AI algorithms are promising for THz communication systems~\cite{chen2021cutting}.

	\item \textbf{Inter- and Intra-path Multiplexing Hybrid Beamforming}.
	In contrast with the lower-frequency MIMO systems that are constrained by the number of antennas, the multiplexing gain of the UM-MIMO system is instead limited by the spatial degree-of-freedom of the THz sparse channel. To this end, an idea to explore additional intra-path multiplexing gain is to enlarge the antenna or subarray spacing, and thereby form the widely spaced multi-subarray (WSMS) architecture~\cite{song2018twolevel,yan2022twolevel}, which is also known as LoS MIMO~\cite{Do2021LoSMIMO}. In the WSMS architecture, the antenna array is composed of multiple subarrays, in which the antenna spacing is half of the wavelength. By contrast, the subarrays are widely-spaced. Each subarray observes the same propagation path with different angles, for which spherical-wave propagation needs to be accounted for. As a result, one propagation path between transmitted and received arrays can be decomposed into multiple sub-paths at the subarray level, which enhances the degree-of-freedom and brings additional intra-path multiplexing gain.
	
	
	\item \textbf{Joint Active and Passive Beamforming}.
	As mentioned in Sec.~\ref{subsec:antennas}, RIS-embedded THz systems have the potential to alleviate the LoS blockage issue as well as to enlarge the coverage, by intelligently controlling the incident wave reflection~\cite{newref1}. Multiple research challenges still remain, including the joint design of active and passive beamforming algorithm in the THz band. On one hand, the active beamforming enabled by antenna array systems takes advantage of the small size and large integration of THz antennas. By concentrating the transmission energy to radiate towards users, the receiving power and transmission coverage can be considerably improved. On the other hand, the RIS is equipped with a metamaterial surface of the integrated circuit, which can be programmed and customized to generate passive beamforming to control the reflection of the incident wave from the transmitting end to the target user and effectively bypass the barrier and increase the selection of transmission routes~\cite{wan2021holographicris,Boulogeorgos2021coverage}.
\item {\textbf{Efficient Baseband Signal Processing}. 
Since the sampling frequency of conventional ADCs and DACs is several tens of Gbps, innovative designs of baseband signal processing are motivated to support the transceivers working with data rates on the order of Tbps~\cite{sarieddeen2021overview}.
Four-folded issues challenge efficient THz baseband signal processing.
First, accurate channel estimation is required to combat the mobility and fast-varying channels to mitigate misalignment, especially with a lack of a LoS path. 
Second, as the largest computational process in communication systems, channel coding should consider novel low-complexity schemes to ensure the low latency and Tbps data rates of THz communications.
Third, efficient data detection methods are needed to address the sophisticated nonlinear inter-channel correlation of THz MIMO systems with low latency.
Last, to further improve the end-to-end time efficiency and optimal resource allocation of the THz baseband signal processing, joint design of coding, modulation, as well as detection can be a promising research direction. }

\end{itemize}
\section{Higher Layer Networking Protocols}
\label{sec_network}

6G and Beyond wireless systems require efficient end-to-end solutions, not only at the physical layer.
The distinct device technologies in Sec.~\ref{sec_device} and attributes of THz waves elaborated in Sec.~\ref{sec:channel char} motivate rethinking the higher layers of the protocol stack.
In this section, we investigate the innovative designs for the data link layer and network layer. 
More specifically, we focus on {MAC} design, interference, and coverage in the data link layer. 
For the network layer, we discuss relaying, routing, and scheduling.

\subsection{Medium Access Control}
The characteristics of THz communications pose the following challenges on MAC design~\cite{han2019medium}.
First, the need for high-gain directional antennas (and, thus, narrow beams) simultaneously at the transmitter and the receiver makes link-layer synchronization extremely challenging due to the deafness problem. First, users need to discover their neighbors, a process that can be highly resource consuming when very narrow beams are used and no separate control channel exists. Second, even when the nodes have situational awareness, they need to be facing each other, which again cannot be easily ensured with traditional coordination solutions. 
Third, due to various human or wall LoS blockages, the cell boundary determined by the received signal strength (RSS) can be not regular circular or hexagonal. Instead, the shape of boundary might become amorphous.

There have been several MAC protocols recently proposed to address these challenges~\cite{relatedAdd5}.
To solve the deafness problem in a more timely and less energy consuming approach than traditional beam search algorithms, we have proposed the use of  receiver-initiated MAC protocols~\cite{synPulseAndCarrier}. With such protocols, users periodically sweep space following a predefined pattern to ensure that all nodes can find each other. The very high data-rates of THz communications then enable ultra-fast data transmission during the communication period for each pair of nodes. 
To accelerate the neighbor discovery, the authors in~\cite{ghasempour2020single} exploit the THz rainbow to obtain the angular information of users.
In particular, the THz rainbow is created by using a leaky waveguide, and the entire angular space is thus illuminated with different THz frequencies.
The neighbor angular location and rotation information are then acquired on the basis of the specific frequency of the received spectrum.
Similarly for network association and neighbor discovery, we have exploited side lobe information of highly THz directional antennas, which is promising to accelerate the neighbor discovery~\cite{macDiscover}.
Moreover, to carefully design the control signaling, we have leveraged the microwave radio  for the exchange of very short control frames to schedule THz transmissions~\cite{yao2016tab,macDualBand}. Specifically, 2.4/5 GHz omnidirectional radio is utilized for control signaling, while THz beamforming is adopted for data transmissions. As a result, THz communication networks are potential to reach the lower delay, higher throughput than existing MAC protocols.
In line with this idea, joint optimization of triple types of radio (THz, mm-wave, and microwave) for concurrent unidirectional data and control signal transmissions is analyzed in~\cite{macTripBand}. 

Recently, we have proposed the use of non-orthogonal multiple access (NOMA) in THz networks as well, with the goals of improving spectral efficiency and promoting fairness across users, especially for the weak user in a network~\cite{macNOMA}. 
The key principles of NOMA are to utilize the power domain and exploit the channel difference of users, in which successive interference cancellation is equipped at the strong user in the NOMA pair.
In light of this, maximization of throughput and energy efficiency can be realized by optimizing the utilization of beam, bandwidth, and power resources with guaranteed fairness~\cite{zhang2020energy,zhang2021energy}. 

To relate MAC with resource allocation, we have proposed an interesting yet unique idea for the THz spectrum in~\cite{Han2014distance}, which exploits the distance-aware and bandwidth-adaptive features and is known as long-user-central-window (LUCW). Specifically, benefited from the distance-adaptive modulations as discussed in~\ref{sec_modulation}, the LUCW principle intelligently allocates the center spectrum of the spectral windows to the long-distance users first, and then the side spectrum to the short-distance users~\cite{han2016distance}.
We further develop this multi-band-based spectrum allocation idea with adaptive sub-band bandwidth by allowing the spectrum of interest to be divided into sub-bands with unequal bandwidths~\cite{shafie2022spectrumallocation}.

\subsection{Interference and Coverage}

The use of high-gain directional antennas at all times requires revisiting some of the well-established multi-user interference and coverage models.
As we first analyzed in~\cite{petrov2017interference}, the interference on a user residing within the main transmitter beam is extremely high, due to the very high antenna directivity gain.
However, the probability of having an interfering user within the main transmit beam is very low. 
We  derive the moment generating functions of the aggregated interference and theoretical expressions for the mean interference powerin~\cite{interferenceCoverageModel2}.
Compared to the RF and mm-wave communications, interference strength in THz directional transmission is much weaker on average. 

On the downside, directional beams lead to a significant beam misalignment problem (as elaborated in Sec.~\ref{sec_beamalignment}) that lessens the coverage and communication quality in THz networks. 
Although dense networks could help coverage enlargement, the expense is the increasing interference as we analyzed in~\cite{interferenceCoverageProblem}.
To this regard, multiple analysis of coverage in THz networks have been conducted.
Based on the aggregated interference and theoretical expressions for the mean interference power, we have derived the approximated coverage probability and average network throughput~\cite{interferenceCoverageModel2}. Numerically, AP density of 0.15/m$^2$, i.e., one AP per 6.7m$^2$, is recommended to support 93\% coverage probability and 30 Gbps/m$^2$ network throughput. 
Apart from the consideration of indoor blockage, we have investigated an accurate coverage model taking into account the 3D property of THz communication~\cite{shafie2021coverage}.
Simulation results demonstrate that increasing the antenna directivity at APs rather than UEs is more efficient to enhance the reliability of THz communications.
By contrast, the accurate coverage analysis with outdoor LoS blockages for THz communications is studied as well~\cite{chen2021coverage}, showing that denser nodes, longer transmission distance, lower latitude and more humid seasons result in lower coverage probability and narrower transmission windows.

Recent research efforts have been made to suppress the interference and enhance the coverage.
As for the interference mitigation, the aforementioned LUCW and the flexible distance-aware scheme~\cite{interferenceCoverageScheme1} can achieve massive connection while preventing interference.
Combined with reinforcement learning, adaptive intermittent interference detection and mitigation in dynamic THz channels are achievable as well~\cite{interferenceCoverageScheme2}.
By contrast, without incurring severe inter-symbol interference, we have illustrated that the multiple access points can remarkably increase the coverage probability~\cite{interferenceCoverageScheme3}.
In particular, with the output power of 1~W, as the number of access points increases from 1 to 20, the coverage probability rises from 25\% to 95\%, respectively.
To further improve the coverage probability, we propose the nearest LoS-AP association that is promising to outperform the traditional nearest AP association~\cite{interferenceCoverageModel2}. 

\subsection{Multi-hop Communications with Active and Passive Relays}
\label{sec_multihop}

Multi-hop scheme is useful to extend the coverage, by using active and passive relays.
On the one hand, by repeating or even amplifying the received signals along transmission paths, relays can combat the distance limitation of the THz propagation.
On the other hand, relays can be leveraged to circumvent the obstacles, forming multiple segments of LoS paths.

Owing to these advantages, we have implemented a multi-hop scheme   to realize the nanoscale THz communication networks~\cite{routing4}. 
In particular, this design enlarges the transmission coverage that is limited by the constrained transmit power in nano-devices.
In macro- or micro-scale THz networks, we have derived  a mathematical model to explore the optimal distances between relays  in~\cite{multihopRelayModel}. More specifically, the effects of THz directional antennas from the perspectives of all the physical, link, and network layers are jointly considered.
Another research direction is on customizing the directions of transmitting for each relay. To this end, the authors in~\cite{multihopRelayBF} derive the closed-form expressions for the THz two-way relay hybrid precoding optimization. 
Furthermore, we illustrate that thanks to the broad THz spectrum resource, efficient transmission among multiple wireless hops enables the integrated access and backhaul (IAB)~\cite{cross2}. Hence, wireless relays in IAB can reduce the cost of wireline connections among base stations, while the IAB topology and signal routing need to be further investigated.
    

In addition to these active relays (e.g. transceivers), we have employed energy-efficient passive beamforming for the intelligent surfaces (detailed in Sec.~\ref{subsec:antennas}) in a multi-hop network~\cite{nie2020beamforming}.
For example, simulation results illustrate the feasibility and advancement of multi-hop THz RIS assisted networks, compared to single-hop systems and those without RIS~\cite{multihopRelayRIS}.
Specifically, the management of direction is supported by deep reinforcement learning (DRL) to achieve the optimization for NP-hard hybrid beamforming.

\subsection{Routing and Scheduling}

    Inherited from previous analysis on multi-hop THz networks in Sec.~\ref{sec_multihop}, efficient routing mechanisms are needed to realize high throughput, low relay, and high reliability, among others.
Indeed, the special THz channel molecular composition and its impact on the available distance-dependent bandwidth need to be considered when devising routing mechanisms.
{For example, owing to the high directivity of THz waves and fast-varying channel, the optimal connection path might change frequently.
Besides, ultra-high data rates challenge buffers and memory storage of THz devices.}
Moreover, with the growth of number of users in the dense network, the formidable optimization problem for routing and scheduling requires novel solutions, such as ML- and AI-enabled methods~\cite{chen2021cutting}. 
The aim is to learn and adapt real-time routing policy with guaranteed high overall network throughput. 
In line with this, we employ tabular reinforcement learning for routing protocol design, by considering the limited buffer size and directional THz signals~\cite{routing3}.
Simulation results demonstrate a low buffer blockage rate and a high packet arrival rate could be achieved.

In nano-scale THz networks with a rather limited buffer size,  we use reinforcement learning to demonstrate its capability to increase the packet delivery rate and lessen the hop count~\cite{routing4}. In the context of intra-body THz communication networks, the molecular adsorption can result in temperature rise and the risk of tissue damage.
Aiming at precluding this effect, by estimating the temperature rise, a temporal correlation-based algorithm is proposed~\cite{routing2}.
As a result, the temperature rise is dramatically mitigated by 75\%-85\% with high energy efficiency.

{Similarly, due to the rather limited memory by considering ultra-data rate transmsision, 
scheduling that determines priority of packet delivery is worth being re-examined for THz networks.}
We have proposed a distributed scheduling algorithm in~\cite{scheduling1} to achieve timely optimal throughput in THz bufferless IoNT.
Particularly, the authors explore the future traffic rate, virtual debts, and channel sensing information captured by each wireless device.
Perpetual communications can be achieved via the joint consideration of both energy consumption and harvesting. Furthermore, we suggest the idling energy should not be neglected, and design a new duty cycle for the receiving nanonodes to reduce the idling energy consumption as efficient scheduling in~\cite{lemic2020idling}.



\subsection{Open Challenges}

\begin{itemize}
	
	\item \textbf{Effective and Efficient Design for Resource Allocation}. 
	In dense THz networks with ultra-large concurrent user demands, joint optimization to ensure {quality of service (QoS)} needs effective and efficient allocation of diverse resources, including time, bandwidth, antennas, power, routes, among others.
	{However, unlike the lower frequency bands, the unique peculiarities of THz waves lead to new challenges. On one hand, the non-linear molecular absorption effect cause non-convexity and NP-hardness, which substantially challenge the resource design if without loosening practical constraints.
	On the other hand, the resources of nano-devices enabled by THz-band communications are limited \ch{with} energy storage, requiring innovative efficient allocation strategies while considering various means of energy harvesting~\cite{lemic2021survey}}. Moreover,
	ML is favored as a candidate to solve such challenges with its powerful learning ability.
	Nevertheless, the high complexity and large storage requirements should be addressed in online training and deployment for practical uses.


	\item \textbf{Transport Layer Protocols}. 
	The wireless multi-Gbps and Tbps links can bring a tremendous boost of the aggregated traffic in  the network.
	This significantly challenges transport layer design regarding congestion control as well as reliable end-to-end transport.
	For example, redesign of the transmission control protocol (TCP) congestion control window mechanism is necessary to handle the traffic dynamics of THz networks. 
	{In addition, the large propagation loss THz waves can incur a higher probability of outage, which thereby leads to possibly higher probabilities of packet loss and re-transmission, challenging the current TCP design.}
	We  demonstrate that under ideal MAC schemes, THz TCP can reach very high data rates~\cite{interferenceCoverageProblem}.
	However, with realistic MAC schemes, the interplay between the contention-based access and the TCP timers is inefficient. 
	Transmissions thereby suffer recurrent timeouts and congestion recovery.
	Motivated by this, further investigation of efficient networks and transport protocols for THz links is needed to guarantee end-to-end reliability.
	
	\item \textbf{Cross Layer Design}.	
	{Ultra-high data rates and super-low end-to-end latency are required by many applications in THz communication networks.
	Hence, simplifying current interactions across protocol stacks becomes significant yet urgent to improve the end-to-end efficiency of network operations.}
	A promising approach is to merge the network and transport layers with the link and even physical layers, designing cross-layer solutions.
	Such solutions, including our previous work~\cite{xia2022crosslayer}, are expected to not only improve the efficiency of the protocol stack, but achieve joint optimization among PHY, MAC, routing, and transport functions as well.
	\end{itemize}

\section{Experimental and Simulation Platforms}
\label{sec_testbeds}
Ten years ago, the research on THz communications was mostly theoretical. Today, thanks to the advancements in the device technologies (Sec.~\ref{sec_device}), THz research is quickly transitioning from theory to practice. In this section, we describe existing THz experimental platforms as well as state-of-the-art simulation and emulation software, which serves as an intermediate step between pure theory and experimental research.

\subsection{Technology Demonstrators and Communication Testbeds}
\label{subsec:testbeds}
We can classify experimental platforms in three categories based on their capabilities: channel sounding systems, technology demonstrators and communication testbeds. Channel sounding systems, which we have already described in Sec.~\ref{channel measurement}, implement no physical layer (in the case of VNA or THz-TDS systems) or a very limited physical layer that supports only the transmission of channel sounding sequences (in the case of time-domain correlators). In this section, we focus on systems that implement at least a physical layer, as listed in {Table~\ref{tab:testbed}} and elaborated as below. Note that if any relevant system is missing, it is not intentional and we apologize ahead of time.

\textbf{Technology demonstrators} are platforms that have been used to experimentally demonstrate the practicality or capabilities of a specific hardware configuration or physical layer solution and, as such, provide limited reconfigurability. 
 A detailed survey shows that lab-level technology demonstrators can be realized in two ways~\cite{chen2019survey}. 
 On one hand, an approach named the solid state THz system generates THz signals by processing the intermediate frequency modulated sources.
 On the other hand,  the high-power THz signal can be directly generated by the direct modulation THz system.

Some of the very early technology demonstrators date back more than ten years, such as the Schottky-diode-based transmission system at 300~GHz first introduced in~\cite{jastrow2008300} and later upgraded in~\cite{jastrow2010wireless}, which supported fixed analog and digital video transmission, respectively. 
{With the bit error rate as low as $1.784\times10^{-10}$, a demonstrator achieved a data rate of 3 Gbps on the basis of super heterodyne Schottky barrier diodes transceivers at 340 GHz over 50~m~\cite{wang20140}.}
Similarly, as a demonstrator of the capabilities of UTC-based photonic transmitters, operating in conjunction with a Schottky-diode-based electronic receiver, a system able to support first 12.5~Gbps and later 24~Gbps at 50~cm was shown in~\cite{song2010terahertz} and~\cite{song201224}, respectively. 
Going up in frequency, a system able to operate at 625~GHz based on again Schottky diodes was shown in~\cite{moeller20112}, with 2.5~Gbps.
The integration of the THz electronic hardware components in a single {Terahertz} Monolithic Integrated Circuit (TMIC) has also enabled the development of more compact THz systems, such as those presented in~\cite{kallfass2011all,antes2012220,koenig2013wireless,kallfass201564,deal2017666,belem2019300}, generally in the sub-300~GHz range (with the exception of~\cite{deal2017666} at 666~GHz), with data-rates of up to 100~Gbps~\cite{koenig2013wireless} and distances of up to 850~m~\cite{kallfass201564}. 
{Recently, a system built by the University of Electronic Science and Technology of China (UESTC) illustrated the ability of Schottky barrier diodes in realizing 20.8 Gbps dual-carrier real-time transmission at 220~GHz with a distance of 1030~m~\cite{feng202120}.
In addition, by operating at 360-430 GHz, Purple Mountain Laboratories established a real-time wireless communication system achieving the data rate of up to 206.25 Gbps~\cite{zhang20226g}.
Furthermore, by combining a $2\times2$ polarized MIMO structure, Huawei Technologies built a prototype that reached 240 Gbps outdoor transmission over a distance of 500~m at 220~GHz with a bandwidth of 13.5~GHz~\cite{huawei20226g}.}

In the recent years, the photonics approach is  becoming more aggressively pursued.
Particularly, up to 106~Gbps links have been realized by a photonic-wireless communication system operating at 400~GHz by Technical University of Denmark and Zhejiang University~\cite{jia20180}. 
{The system was further upgraded recently to reach 131~Gbps single-channel data rate with $-24$~dBm emitted power~\cite{jia2022integrated}.}
Designed for a $2\times2$ MIMO link, another photonics-based THz system has demonstrated a $6\times 20$~Gbps data rate as well, with the frequency ranging from 375~GHz to 500~GHz in six channels~\cite{li2019120}.
In addition to the photonics-aided transmitter, by deploying analog down-conversion through an electrical mixer at the receiver side and electrical-to-optical conversion via direct modulation, up to 13~Gbps fiber-THz-fiber communication has been realized at 450 GHz~\cite{wang2018fiber}.

Customized solutions for specific applications have also been presented. For example, the first functional demonstrator for THz kiosks, aimed at transferring a large amount of data from a stationary terminal to a mobile wireless device in a very short time under the ``touch-and-go'' scenario, has recently been demonstrated in~\cite{testbedKIOSK}. 
In this platform, InP-based front-ends at 300 GHz are utilized in conjunction with a high-speed field programmable gate array (FPGA)-based physical layer that implements Reed-Solomon (RS) FEC to demonstrate 20 Gbps of real-time data.
Another specific application, in which THz communications play a key role, is  uncompressed high-definition (HD) (60 fps) and 4K (30 fps) video transmission. In~\cite{testbedLiveStream}, a UTC-based photonic transmitter is utilized in conjunction with a Schottky-diode-based receiver at 138 GHz to demonstrate video transmission over 30~cm with 99\% of undamaged frames in HD and 95\% in 4K. 

The goal of \textbf{communication testbeds} is to facilitate the testing of new communication solutions for ultrabroadband networks in the THz band. As such, a testbed should be highly reconfigurable and easy to reprogram to support different hardware blocks, signal processing algorithms, or communication and networking protocols. In this direction, in~\cite{sen2019experimental,sen2020teranova,sen2021versatile} we introduced the \textit{TeraNova} platform, the world’s first integrated testbed for ultrabroadband communication networks in the THz band. The highly modular testbed consists of several analog up \& down converting front-ends at 120-140 GHz, 210-240 GHz and 1-1.05 THz, making it not only the highest frequency platform existing to date, but also multi-band. The front-ends can be interfaced to different available DSP back-ends able to process baseband bandwidths ranging from 2~GHz in real-time and up to 32~GHz offline, per channel. Directional antennas with gains ranging from 21~dBi to 55~dBi are available at the different frequencies. The TeraNova platform supports the development of 6G THz systems in many ways, including but not limited to: 1) ultra-broadband channel sounding in indoor and outdoor scenarios; 2) design of ultra-broadband synchronization, channel estimation and equalization, modulation and coding; and, 3) testing of directional medium access control, neighbor discovery, and multi-hop relaying. Remote access is also supported. Some of the early data-sets collected with the testbed are available in SigMF format~\cite{teranova}.

Major equipment vendors have also recently entered the arena above 100~GHz. In~\cite{ni2019wp}, National Instruments (NI) exemplifies how their mm-wave transmitter systems (MTS) can be extended to operate above 100~GHz by directly connecting their FPGA-based baseband to THz up \& down converters. The MTS supports different functionalities, including channel sounding as well as pre-defined TDMA-type and OFDM-type physical layers. In~\cite{keysight2020wp}, Keysight Technologies demonstrate how their commercial signal generators, arbitrary waveform generators, signal analyzers and digital storage oscilloscopes can be used for testing of hardware components, channel models and signal processing algorithms, when combined with up \& down converters above 100~GHz. Interestingly, both vendors as well as many of the aforementioned electronic-based THz systems rely on frequency up \& down converters built by Virginia Diodes Inc.~\cite{vdi_updown}.

\begin{table*}
\scriptsize
\centering
		\caption{Technology demonstrators and communication testbeds for THz communications.}
	\label{tab:testbed}
\begin{tabular}{|p{2cm}<{\centering}|p{4cm}<{\centering}|c|p{2cm}<{\centering}|p{5.5cm}<{\centering}|}
\hline
\textbf{Type}                             & \textbf{Title and Reference}                                                                                                                         & \textbf{Year} & \textbf{Affiliation}                                                                    & \textbf{Achievement}                                                                                                                                                                              \\ \hline
\multirow{16}{2cm}{\centering Technology demonstrator} & Wireless Digital Data Transmission at 300~GHz~\cite{jastrow2010wireless}                                                                             & 2010          & Physikalisch-Technische Bundesanstalt                                                   & Schottky-diode-based fixed digital video transmission system at 300 GHz                                                                                                                           \\ \cline{2-5} 
                                          & 2.5 Gbit/s Duobinary Signalling with Narrow Bandwidth 0.625 Terahertz Source~\cite{moeller20112}                                                     & 2011          & Bell Laboratories/ Alcatel-Lucent                                                       & Schottky-diode-based system at 625 GHz                                                                                                                                                            \\ \cline{2-5} 
                                          & 24~Gbit/s Data Transmission in 300~GHz Band for Future Terahertz Communications~\cite{song201224}                                                    & 2012          & Nippon Telegraph and Telephone (NTT) Laboratories                                       & UTC-based photonic transmitter and Schottky-diode-based electronic receiver supporting 24 Gbps                                                                                                    \\ \cline{2-5} 
                                          & Wireless Sub-THz Communication System with High Data Rate~\cite{koenig2013wireless}                                                                  & 2013          & Karlsruhe Institute of Technology                                                       & Terahertz Monolithic Integrated Circuit (TMIC) in the sub-300~GHz range with data-rates of up to 100~Gbps                                                                                         \\ \cline{2-5} 
                                          & 0.34-THz Wireless Link Based on High-Order Modulation for Future Wireless Local Area Network Applications~\cite{wang20140}                           & 2014          & China Academy of Engineering Physics                                                    & 3 Gbps super heterodyne Schottky-diode-based transmission system at 340 GHz with the bit error rate of $1.784\times10^{-10}$                                                                      \\ \cline{2-5} 
                                          & 64~Gbit/s Transmission over 850~m Fixed Wireless Link at 240~GHz Carrier Frequency~\cite{kallfass201564}                                             & 2015          & Fraunhofer Institute for Applied Solid-State Physics/ Karlsruhe Institute of Technology & TMIC in the sub-300~GHz range with distances of up to 850~m                                                                                                                                       \\ \cline{2-5} 
                                          & A 666~GHz Demonstration Crosslink with 9.5~Gbps Data Rate~\cite{deal2017666}                                                                         & 2017          & Northrop Grumman Corporation                                                            & TMIC at 666~GHz                                                                                                                                                                                   \\ \cline{2-5} 
                                          & Fiber-THz-Fiber Link for THz Signal Transmission~\cite{wang2018fiber}                                                                                & 2018          & Fudan University                                                                        & 13~Gbps fiber-THz-fiber communication at 450~GHz with photonics-aided transmitter and electrical mixer based receiver                                                                             \\ \cline{2-5} 
                                          & Prototype of KIOSK Data Downloading System at 300 GHz: Design, Technical Feasibility, and Results~\cite{testbedKIOSK}                                & 2018          & Pohang University of Science and Technology                                             & 20 Gbps of real-time data with InP-based front-ends at 300~GHz                                                                                                                                    \\ \cline{2-5} 
                                          & Live Streaming of Uncompressed HD and 4K Videos Using Terahertz Wireless Links~\cite{testbedLiveStream}                                              & 2018          & Polytechnique Montr\'{e}a                                                               & 99\% of undamaged frames in HD and 95\% in 4K videos at 138 GHz                                                                                                                                   \\ \cline{2-5} 
                                          & 300~GHz Quadrature Phase Shift Keying and QAM16 56~Gbps Wireless Data Links Using Silicon Photonics Photodiodes~\cite{belem2019300}                  & 2019          & University of Lille                                                                     & TMIC in the sub-300~GHz range                                                                                                                                                                     \\ \cline{2-5} 
                                          & 120 Gb/s Wireless Terahertz-Wave Signal Delivery by 375 GHz-500 GHz Multi-Carrier in a 2$\times$ 2 MIMO System~\cite{li2019120}                      & 2019          & Fudan University                                                                        & $6\times 20$~Gbps photonics-based system data rate at 375-500~GHz                                                                                                                                 \\ \cline{2-5} 
                                          & A 20.8-Gbps Dual-Carrier Wireless Communication Link in 220-GHz Band~\cite{feng202120}                                                               & 2021          & University of Electronic Science and Technology of China                                & 20.8 Gpbs dual-carrier Schottky-diode-based real-time transmission system at 220 GHz with a distance of 1030~m                                                                                    \\ \cline{2-5} 
                                          & 6G Oriented 100 GbE Real-time Demonstration of Fiber-THz-Fiber Seamless Communication Enabled by Photonics~\cite{zhang20226g}                        & 2022          & Purple Mountain Laboratories                                                            & 206.25 Gbps real-time wireless communication system at 360-430 GHz                                                                                                                                \\ \cline{2-5} 
                                          & Integrated Dual-Laser Photonic Chip for High-Purity Carrier Generation Enabling Ultrafast Terahertz Wireless Communications~\cite{jia2022integrated} & 2022          & Technical University of Denmark and Zhejiang University                                                        & 131~Gbps photonic-wireless communication system at 400~GHz with -24 dBm power                                                                                                                     \\ \cline{2-5} 
                                          & 6G ISAC-THz Opens up New Possibilities for Wireless Communication Systems~\cite{huawei20226g}                                                        & 2022          & Huawei Technologies                                                                     & 240~Gbps outdoor transmission at 220~GHz with bandwidth of 13.5~GHz over 500~m                                                                                                     \\ \hline
\multirow{5}{2cm}{\centering Communication testbed}    & Creating a Sub-Terahertz Testbed with the NI mmWave Transceiver System~\cite{ni2019wp}                                                               & 2019          & National Instruments                                                                    & Operate above 100~GHz by directly connecting FPGA-based baseband to THz up \& down converters, support channel sounding and pre-defined TDMA-type/OFDM-type physical layers                       \\ \cline{2-5} 
                                          & Experimental Demonstration of Ultra-Broadband Wireless Communications at True Terahertz Frequencies~\cite{sen2019experimental}                       & 2019          & \multirow{2}{2cm}{\centering Northeastern University}                                                & \multirow{2}{5.5cm}{\centering First integrated testbed for THz communication networks covering 120-140 GHz, 210-240 GHz and 1-1.05 THz simultaneously, PHY testing with up to 32-GHz of bandwidth per channel.} \\ \cline{2-3}
                                          & The TeraNova Platform: An Integrated Testbed for Ultra-Broadband Wireless Communications at True Terahertz Frequencies~\cite{sen2020teranova}        & 2020          &                                                                                         &                                                                                                                                                                                                   \\ \cline{2-5} 
                                          & A New Sub-Terahertz Testbed for 6G Research~\cite{keysight2020wp}                                                                                    & 2020          & Keysight                                                                                & Testing of hardware components, channel models, and signal processing algorithms with up \& down converters above 100~GHz                                                                         \\ \cline{2-5} 
                                          & A Versatile Experimental Testbed for Ultrabroadband Communication Networks above 100~GHz~\cite{sen2021versatile}                                     & 2021          & Northeastern University                                                                 & First ultra-broadband spread spectrum reaching above 1 THz, MAC and above testing with 8~GHz of real-time bandwidth                                                                               \\ \hline
                                          \end{tabular}	
\end{table*}

\subsection{Simulation Tools}
\label{subsec:simulators}
In parallel to the development of experimental testbeds, simulation tools are needed to expedite the development of communication and networking protocols tailored to THz systems, at a fraction of the cost. Simulation platforms can be categorized into propagation/channel simulators and communication/networking simulators.

\textbf{Channel simulators} are aimed at reducing the costly and time-demanding process of collecting extensive data sets for different scenarios with changing geometry, weather conditions, etc.
With generalized parameters extracted from measurements, channel simulators can virtually approximate wireless propagation models on demand.
Different options are available to date. Commercial software, such as EDX Advanced Propagation and Wireless InSite can be utilized at frequencies up to 0.1 THz~\cite{modelSimCloud}.
Academia-driven platforms, such as \textit{CloudRT} developed by BJTU and Technische Universit\"at Braunschweig, supports the simulation of both urban indoor and outdoor cases at 450~MHz-325~GHz~\cite{modelSimCloud}.
Ad-hoc channel solvers that build on experimentally-measured data have been utilized to explore the performance of different specific scenarios, including THz backhaul links at 300 GHz~\cite{modelSim300G} or multimedia kiosks~\cite{modelAddKiosk}. While many of the existing platforms are for single channel simulators, \textit{TeraMIMO}~\cite{tarboush2021teramimo} has been recently developed to simulate UM MIMO channels in 3D scenarios, capturing the properties described in Sec.~\ref{sec_PHY}. Other simulation platforms, such as~\cite{thorSimulator}, are being developed with the aim of not only simulating the channel, but also testing certain physical layer solutions.


While at a lower pace than channel simulators, THz \textbf{network simulators} are also being developed. THz network simulation tools are designed with the objective of testing, refining and optimizing networking protocols for THz networks, such as those described and envisioned in Sec.~\ref{sec_network}. Aimed at the simulation of nanoscale THz communication networks, \textit{Nano-Sim} was first introduced in~\cite{piro2013nano}. Nano-sim is built as an extension to \textit{ns-3}, which is one of the most widely used teaching and education network simulation software, and incorporates a simplified channel model for ultra-short range THz communications and a tailored protocol stack. More recently, in~\cite{hossain2018terasim}, we introduced \textit{TeraSim}, an open source network simulation platform for THz communication networks that is also built as an ns-3 extension and it is officially available in the ns-3 App Store. The simulator has been developed considering two major types of application scenarios, namely, nanoscale communication networks and macroscale communication networks. The simulator consists of a common channel module that accurately captures the physics of the THz channel, separate physical and link layers tailored to each scenario, and two assisting modules, namely, THz directional antenna module and energy harvesting module, originally designed for the macroscale and nanoscale scenario, respectively. Terasim seamlessly interfaces with existing ns-3 modules, including higher layer protocols and mobility models. Since its release in 2018, different groups have adopted TeraSim, and extended new capabilities, such as the inclusion of the latest experimentally-validated multi-path THz channel models~\cite{gargari2021full}.

{\subsection{Open Challenges}
These are the main challenges impacting the development of experimental and simulation platforms:}
\begin{itemize}
    \item {\textbf{Real-time Experimental Platforms for THz Networking:} Most of the experimental platforms described in Sec.~\ref{subsec:testbeds} rely on high-performance arbitrary waveform generators in charge of generating previously-designed ultra-broadband baseband or intermediate frequency signals at the transmitters and, reciprocally, high-performance digital storage oscilloscopes that can digitize and save for later processing the received down-converted signals. Such approach is excellently suited for channel sounding and testing physical-layer solutions, but do not support the testing of networking protocols as signals are not processed in real-time. The design of a custom integrated circuit to implement a pre-defined physical layer would solve this problem, but as discussed in Sec.~\ref{sec_PHY}, the physical layer of THz communication systems is still being designed. Alternative, programmable circuits such as FPGAs need to be utilized. The NI testbed~\cite{ni2019wp} implements a 5G NR-like physical layer on an FPGA and, thus, operates in real-time, but only with 2 GHz of channel bandwidth. In our vision, radio-frequency systems on chip (RFSoC), which integrate high-performance FPGAs with large numbers of DACs and ADCs can be leveraged to design new experimental platforms with larger real-time bandwidth. Early steps in this direction has been reported recently in~\cite{abdellatif2022demo}}.
    
    \item {\textbf{Modular, Upgradable Hardware Architectures:} The majority of existing testbeds consist of very specific digital back-end signal processors, analog front-ends and antenna systems, difficult to upgrade if not as a whole. The definition of standard physical and logical plug-and-play interfaces to interconnect the different testbed building blocks would drastically facilitate the upgrade of specific building blocks as better solutions become available. While this is not the case for user-equipment devices, where the highest system performance is obtained by integrating all components in a single chip, it is very important for platforms that are designed to test new solutions. Standardized physical interfaces for analog components exist, such as rectangular waveguide (WR) connectors for different frequency bands, or even coaxial cables able to support different bandwidths, but for example a logical interface designed for the high-speed transfer of I\&Q samples between a digital back-end and the analog front-end is missing. In this direction, some of the principles of O-RAN~\cite{bonati2020open}, aimed at decoupling the different hardware and software blocks of future cellular networks to ensure inter-operability between vendors, could serve as inpiration.}
    
    \item {\textbf{Integration of Hardware and Software Platforms through Large-scale Emulation:} The cost of THz experimental platforms limit the possibility to test the performance of large THz networks. Simulation platforms instead rely on abstractions or mathematical models of the device components and the wireless channels to then test networking protocols. Half-way between the two, emulation platforms with harwdare in the loop can be developed to better capture the real behavior of THz devices and channel. For example, Colosseum, i.e., the largest RF emulator in the world~\cite{bonati2021colosseum}, has been utilized to test and refine networking solutions for massive networks with large number of devices spanning up to 1~km$^2$. However, the current Colosseum cannot emulate omnidirectional emitters at frequencies under 6~GHz. The expansion of the existing platform or the development of a new emulator able to cover higher frequencies and directional transmissions will facilitate the testing of high-density THz networks.}
\end{itemize}

\section{Policy and Standardization}
\label{sec_standards}
\subsection{Spectrum Policy}

As we discussed in Sec.~\ref{sec_channel}, the THz channel provides very large bandwidths, which through new device technologies (Sec.~\ref{sec_device}) and innovative communication (Sec.~\ref{sec_PHY}) and networking solutions (Sec.~\ref{sec_network}) will enable transformative applications. However, current spectrum policy legally limits the potential of the THz band. More specifically, as briefly mentioned before, THz frequencies are of interest not only to the commercial, industrial and military communication sectors, but also to the scientific sensing community. The same molecular absorption that impacts THz communications enables the sensing of critical atmospheric information relating to weather forecasting~\cite{suen2014global} and even climate change monitoring~\cite{thakshila2021climate}. Such sensing systems are either based on ground-based radio-telescopes or, primarily, in satellites orbiting the Earth as part of Earth Exploration Satellite Services (EESS). 

Earth atmospheric and planetary sensing signals resulting from molecular processes are extremely weak and can be very easily impacted by interference from communication systems. Because of this, several frequency bands in the electromagnetic spectrum are protected, i.e., no human-made radiation is allowed. Such ``prohibited" bands are defined by the International Telecommunication Union (ITU) Radio Regulation (RR) 5.340 and, thus, have worldwide impact. While these bands exist across the spectrum, their presence and width is larger at frequencies above 100~GHz. Particularly, these are the forbidden bands: 100--102~GHz, 109.5--111.8~GHz, 114.25--116~GHz, 148.5--151.5~GHz, 164--167~GHz, 182--185~GHz, 190--191.8~GHz, 200--209~GHz, 226--231.5~GHz, 250--252~GHz. Moreover, neighboring with these frequencies, there are other passive bands in which spectrum sharing between communication and sensing services is very limited. 

Resulting from these, as of today, the largest contiguous bandwidth  between 100 and 200~GHz that can be legally utilized is 12.5~GHz, i.e., much less than what the physics of the channel establish. To be able to utilize larger bandwidths, one needs to move closer to the 300~GHz range. While on paper this is a ``minor shift in frequency", technologically speaking, the technology capabilities at 300~GHz in terms of transmission power, noise figure or conversion losses, are much lower than those at 100~GHz, and the situation just gets worse as we go up in frequency, as per the discussion in Sec.~\ref{sec_device}. Therefore, rather than ``giving up" on the frequencies between 100~GHz and 300~GHz, developing new methods for coexistence between commercial and military communication systems and scientific sensing services is of critical importance. 

Interestingly, the origin of ITU RR 5.340 dates back to the ITU World Radio Conference in the year 2000. WRC 2000 Resolution 731 defined such bands and,  at the same time, stated that sharing between active services (communications and radar) and passive (sensing) should be a goal if the technical means can be found to enable it. Moreover, those means should not be solely based on communication technologies being adapted around sensing services, but should take into account the principles of burden-sharing to the extent practicable, i.e., both active and passive services should consider modifying their long-term designs to facilitate sharing.

As discussed at length in~\cite{polese2021coexistence}, there are many potential approaches to facilitate coexistence and spectrum sharing between passive and active users of the THz band, which span the entire protocol stack. First, due to the fact that the majority of passive services are in EESS, intelligent radiating structures (ranging from inverted pyramid antennas to frequency-selective surfaces) can be utilized to reduce the emissions in high elevation angles. This is practical for traditional networking architectures in which base stations or access points provide service to ground users, but becomes more challenging as non-terrestrial-networks (e.g., drones, balloons, satellites or any other high-altitude radio platforms) enter the game in 6G~\cite{giordani2020non}. More dynamic solutions, such as dynamic beamforming that takes into account an additional constraint for the high elevation angles only when EESS are in the vicinity can be implemented through ultra-massive MIMO (Sec.~\ref{DAoSA}). At the physical layer, waveforms that can facilitate coexistence can be leveraged. For example, in~\cite{bosso2021ultrabroadband}, we designed, implemented and experimentally tested an ultra-broadband direct sequence spread spectrum (DSSS) system precisely to ensure coexistence between ground-servicing backhaul infrastructure (which have much higher transmission power than user-equipment) and EESS. In terms of networking, dual-band systems that monitor and track EESS and accordingly allocate resources in different frequency bands can be developed too. In this direction, we have recently built, programmed and demonstrated real-time data-carrying system able to dynamically switch between the 120--140~GHz and the 210--240~GHz band according to the presence of a specific EESS (i.e., the NASA AURA satellite)~\cite{polese2022non}.

Besides technical work, and in order to be able to test such solutions, several ongoing legal efforts are happening in parallel. On the one hand, through the Spectrum Innovation Initiative (SII), the National Science Foundation (NSF) has established a program that will culminate in the creation of one or more National Radio Dynamic Zones (NRDZs), or areas in which, under the premise that no radiation escapes the region, the only limitations for the use of spectrum are the laws of physics. Among others, such NRDZs are going to be the playground to explore and test new solutions to exist coexistence. Still within the limits of current spectrum policy, the US Federal Communications Commission (FCC) has also established the concept of Innovation Zones (IZs), or areas in which testing can be performed without the need of requesting an experimental license. While there are four IZs in the USA, only the one hosted at Northeastern University currently supports testing at several frequencies above 100~GHz (and in which many of the aforementioned experimental works were conducted). Moreover, since February 2022, the US FCC and US National Telecommunications and Information Administration (NTIA) have commenced a new initiative to improve coordination on spectrum management~\cite{fcc_ntia2022}. The joint efforts by these two government agencies will facilitate a national spectrum strategy for long-term planning and coordination.

\subsection{Standardization}
The standardization activities are on the way since the last several years.
The first wireless communications standard, IEEE 802.15.3d (WPAN), was published in 2017, which operates at the 300~GHz frequency range to support 100 Gbps and above wireless point-to-point links~\cite{IEEE802_15_3d}.
In November 2019, ITU-R WRC-19 Agenda Item 1.15 identified in total 137~GHz bandwidth for land mobile and fixed services, covering the spectrum bands over 0.275--0.296~THz, 0.306--0.313~THz, 0.318--0.333~THz, and 0.356--0.45~THz~\cite{kurner2020impact}.

However, the document provides limited channel models in specific scenarios, centered around 300~GHz only. To accelerate the standardization of the THz band,  application-specific studies of THz channels over the full THz band are still greatly demanded. By reviewing the timeline of 5G standardization process, the channel model standardization for 6G possibly begins in 2023-2025, expanding the supporting frequency to the THz band. There are various expectations on THz channel models in the future 6G standardization, including specification of new application scenarios for THz communications, coverage of new system components such as UM-MIMO and RIS in Sec.~\ref{subsec:antennas}, and emerging methods such as hybrid of deterministic and statistical approaches in Sec.~\ref{sec:SISO-hybrid}.

More recently, a new ITU-R Report entitled ``IMT Above 100~GHz'' was started at the August 2021 meeting of ITU-R WP5D, to study the technical feasibility of IMT in bands above 100~GHz~\cite{imt2021above100ghz}. IMT technologies adopted for bands above 100~GHz can be used in indoor/outdoor hotspot environments, integrated sensing and communication and ultra-short-range environments to provide ultra-high data rate services.
In July 2021, the IEEE Communications Society (ComSoc) Radio Communications Committee (RCC) Special Interest Group (SIG) on THz communications has been established, with the objectives and missions to advance the research and development of THz communications in 6G and Beyond~\cite{thzsig}.

\section{Conclusions}
THz communications are envisioned as a key technology for 6G and Beyond. The very large available bandwidth at THz frequencies (tens of GHz and more, i.e., two orders of magnitude more than that in the 5G systems) will drastically improve the performance of common network applications enabling Tera-WiFi, Tera-IoT, Tera-IAB; whereas the very small wavelength of THz waves opens the door to the new world of nanoscale electromagnetic communications networks such as IoNT, and WiNoC. Furthermore, by integrating the functions of sensing and communication, the THz ISAC applications include THz VR/AR, and vehicular communication and radar sensing.

Building on over ten years of research, in this paper, we have surveyed the recent advancements and highlighted the open research directions of THz communications, with the following take-away lessons.
First, the THz technology gap is almost filled, thanks to the recent progress on THz device technologies, in particular that made in the last decade. However, still meaningful challenges remain related to both analog RF \& digital hardware as well as integrated circuits at THz frequencies.
Second, the propagation and channel modeling attract much attention to discover the spectrum peculiarities. Although recent efforts in indoor channel measurement at 140~GHz, 220~GHz, 300~GHz are reported, extensive studies are still needed at higher frequencies in various indoor and outdoor environment as well as less traditional realms, including intra-body or space. 
Third, in terms of communications and signal processing, major challenges are to combat the distance limitation, reduce power consumption and support mobility coverage, by revisiting and inventing new techniques to make the most use of the broad bandwidth.
Fourth, in THz networks with narrow beams, bandwidth is indeed no longer a constraint. Nevertheless, challenges are faced in medium access control, interference and coverage, multi-hop communications, routing and scheduling.
Fifth, THz band interests not only to communications, but also to other scientific communities including sensing, imaging, astronomy, among others. Therefore, policy and regulation need to be made for efficient coexistence. Finally, although there is a standard for point-to-point links, great efforts and competitions are foreseen in the next a few years in 3GPP and ITU on THz communication standardization.





\bibliography{References}
\bibliographystyle{IEEEtran}

\begin{IEEEbiography}
[{\includegraphics[width=1in,height=1.25in,clip,keepaspectratio]{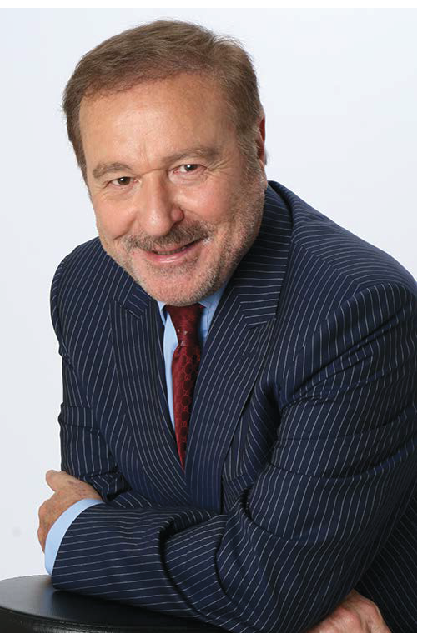}}] 
{\bf I. F. Akyildiz} (Life Fellow, IEEE) received the B.S., M.S., and Ph.D. degrees in Electrical and Computer Engineering from the University of Erlangen–N{\"u}rnberg, Germany, in 1978, 1981, and 1984, respectively. He is  Founder and President of the Truva Inc., a consulting company based in Georgia, USA,  since 1989.  He is an Adjunct Professor with University of Iceland (since 2020), University of Helsinki (since 2021), University  of Cyprus (2017). He is also an Advisory Board member at the Technology Innovation Institute (TII) Abu Dhabi, United Arab Emirates, since June 2020. He is the Founder and the Editor-in-Chief of the newly established of  International  Telecommunication Union Journal on Future and Evolving Technologies (ITU J-FET) since August 2020. He served as the Ken Byers Chair Professor in Telecommunications, the Past Chair of the Telecom Group at the  ECE, and the Director of the Broadband Wireless Networking Laboratory, Georgia Institute of Technology, from 1985 to 2020.
He had many international affiliations during his career and   established  research centers in Spain, South Africa, Finland, Saudi Arabia, Germany, Russia, India. Dr. Akyildiz is an IEEE Fellow since 1996, and ACM Fellow since 1997. He received numerous awards from IEEE, ACM, and other professional organizations, including Humboldt Award from Germany and Tubitak Award from Turkey. In March 2022, according to Google Scholar his h-index is 132 and the total number of citations to his articles is more than 134+K. His current research interests include TeraHertz Band Communciation, 6G/7G wireless systems, Reconfigurable Intelligent Surfaces,Hologram communication, Extended Reality Wireless Communication,  Internet of Space Things/CUBESATs, Internet of Bio-Nano Things, Molecular Communication, Underwater and Underground Communications.
\end{IEEEbiography}

\begin{IEEEbiography}
[{\includegraphics[width=1in,height=1.25in,clip,keepaspectratio]{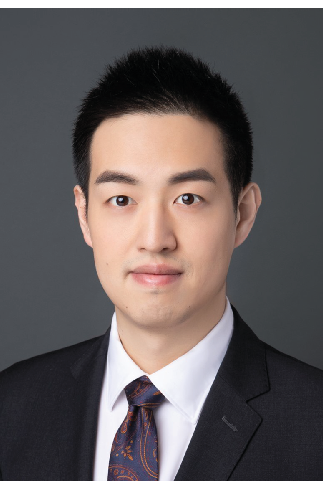}}] 
{\bf Chong Han} (Member, IEEE) received Ph.D. degree in Electrical and Computer Engineering from Georgia Institute of Technology, USA in 2016. He is currently an Associate Professor with the Terahertz Wireless Communications (TWC) Laboratory, University of Michigan–Shanghai Jiao Tong University (UM-SJTU) Joint Institute, Shanghai Jiao Tong University, China. Since 2021, he is also affiliated with Department of Electronic Engineering, Shanghai Jiao Tong University. He is the recipient of 2018 Elsevier NanoComNet (Nano Communication Network Journal) Young Investigator Award, 2017 Shanghai Sailing Program 2017, and 2018 Shanghai ChenGuang Program. He is a guest editor with IEEE Journal on Selected Topics in Signal Processing (JSTSP) and IEEE Transactions on Nanotechnology, an editor with IEEE Open Journal of Vehicular Technology since 2020, IEEE Access since 2017, Elsevier Nano Communication Network journal since 2016. He is a TPC chair to organize multiple IEEE and ACM conferences and workshops. His research interests include Terahertz band and millimeter-wave communication networks, and intelligent sustainable power systems. He is a member of the IEEE and ACM.
\end{IEEEbiography}

\begin{IEEEbiography}
[{\includegraphics[width=1in,height=1.25in,clip,keepaspectratio]{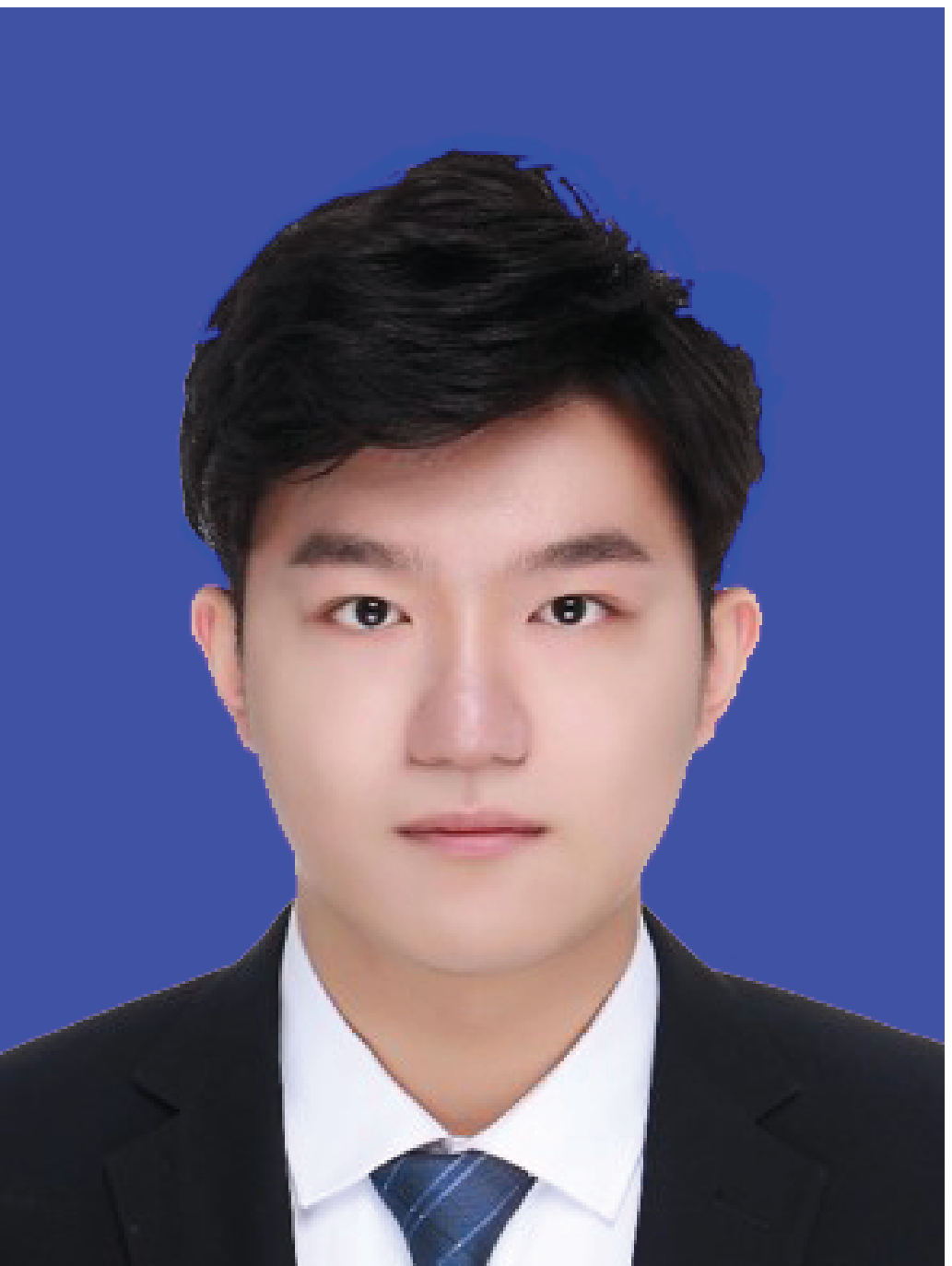}}] 
{\bf Zhifeng Hu} received the B.E. degree in electrical and computer engineering from the Shanghai Jiao Tong University, China, in 2020. He is currently pursuing the Ph.D. degree with the Terahertz Wireless Communications (TWC) Laboratory, University of Michigan–Shanghai Jiao Tong University (UM-SJTU) Joint Institute, Shanghai Jiao Tong University, China. His current research interests include machine/deep learning for Terahertz networking.
\end{IEEEbiography}

\begin{IEEEbiography}
[{\includegraphics[width=1in,height=1.25in,clip,keepaspectratio]{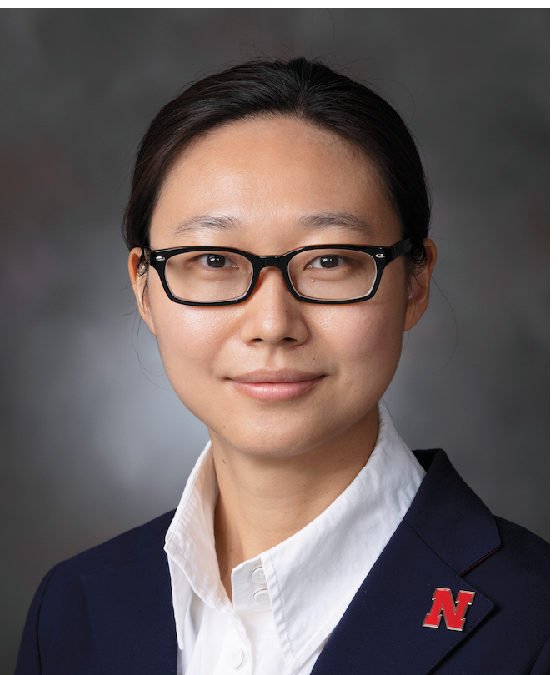}}] 
{\bf Shuai Nie} (Member, IEEE) received the B.S. degree in Telecommunications Engineering from Xidian University in 2012, the M.S. degree in Electrical Engineering from New York University in 2014, and the Ph.D. degree in Electrical and Computer Engineering from Georgia Institute of Technology in 2021. She has joined the School of Computing at University of Nebraska-Lincoln as an assistant professor since August 2021. Her research interests include millimeter wave and terahertz band communication systems and networks.
\end{IEEEbiography}

\begin{IEEEbiography}
[{\includegraphics[width=1in,height=1.25in,clip,keepaspectratio]{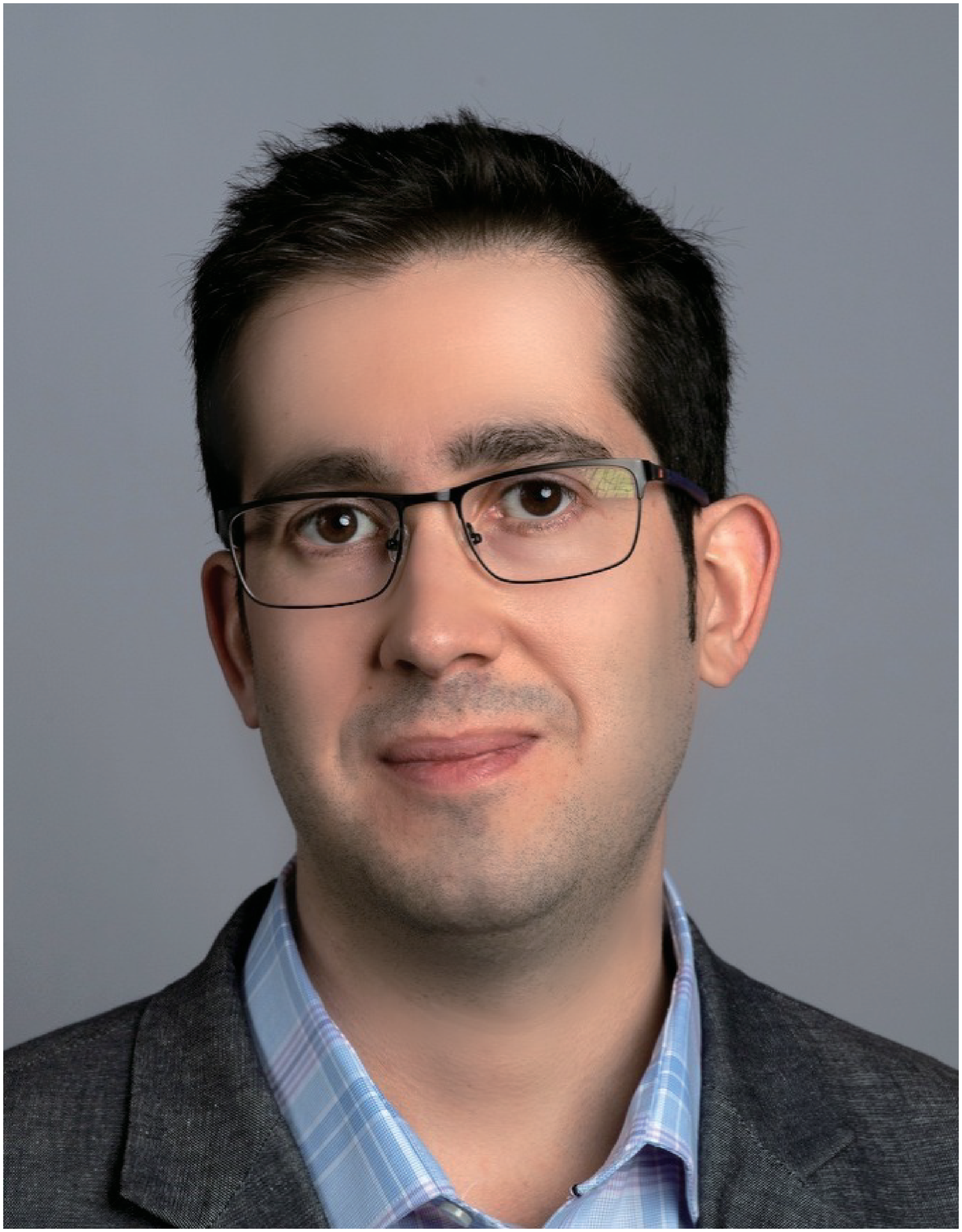}}] 
{\bf Josep Miquel Jornet} (Senior Member, IEEE) received the B.S. in Telecommunication Engineering and the M.Sc. in Information and Communication Technologies from the Universitat Polit\`{e}cnica de Catalunya, Barcelona, Spain, in 2008. He received the Ph.D. degree in Electrical and Computer Engineering from the Georgia Institute of Technology (Georgia Tech), Atlanta, GA, in 2013. 
Between August 2013 and August 2019, he was a faculty with the Department of Electrical Engineering at the University at Buffalo (UB), The State University of New York. Since August 2019, he is an Associate Professor in the Department of Electrical and Computer Engineering, the Director of the Ultrabroadband Nanonetworking Laboratory, and a Member of the Institute for the Wireless Internet of Things at Northeastern University (NU), in Boston, MA. His research interests are in terahertz communication networks, wireless nano-bio-communication networks and the Internet of Nano-Things. In these areas, he has co-authored more than 180 peer-reviewed scientific publications, 1 book, and has also been granted 4 US patents. Since July 2016, he is the Editor-in-Chief of the Nano Communication Networks (Elsevier) Journal. He is serving as the lead PI on multiple grants from U.S. federal agencies including the National Science Foundation, the Air Force Office of Scientific Research and the Air Force Research Laboratory. He is a recipient of the National Science Foundation CAREER award and of several other awards from IEEE, ACM, UB and NU.
\end{IEEEbiography}

\end{document}